\documentclass[sigconf,screen,pbalance]{acmart}
\AtBeginDocument{%
  }

\setcopyright{cc}
\setcctype{by}
\acmDOI{10.1145/3832783.3834369}
\acmYear{2026}
\copyrightyear{2026}
\acmISBN{979-8-4007-2882-2/2026/10}
\acmConference[ASE '26]{Proceedings of the 41st IEEE/ACM International Conference on Automated Software Engineering}{October 12--16, 2026}{Munich, Germany}
\acmBooktitle{Proceedings of the 41st IEEE/ACM International Conference on Automated Software Engineering (ASE '26), October 12--16, 2026, Munich, Germany}
\acmSubmissionID{ase26main-p874-p}
\received{2026-03-26}
\received[accepted]{2026-06-18}

\usepackage{framed}
\usepackage{listings}
\usepackage{xcolor}
\usepackage{color}
\usepackage{multirow}
\usepackage{algorithm}

\usepackage{amsmath}
\usepackage{amssymb}
\usepackage{threeparttable}
\usepackage{url}
\usepackage[most]{tcolorbox}
\usepackage{hyperref}
\usepackage[hyphenbreaks]{breakurl}
\usepackage{algpseudocode}
\usepackage{colortbl}
\usepackage{booktabs}
\usepackage{pbalance}
\usepackage{xkeyval}
\hypersetup{
    colorlinks = true,
    linkcolor=blue,
    filecolor=blue,      
    urlcolor=blue,
    citecolor=cyan,
}
\usepackage{xspace}
\usepackage{subfigure}
\usepackage{graphicx}
\usepackage{fvextra}
\usepackage{pifont}

\usepackage[shortlabels]{enumitem}
\setlist[enumerate]{nosep}

\definecolor{codegreen}{rgb}{0,0.6,0}
\definecolor{codegray}{rgb}{0.5,0.5,0.5}
\definecolor{codepurple}{rgb}{0.58,0,0.82}
\definecolor{backcolour}{rgb}{0.95,0.95,0.92}
\definecolor{lightgreen}{rgb}{0,0.4,0}
\definecolor{lightred}{rgb}{0.4,0,0}

\definecolor{mygray}{gray}{.9}

\lstdefinestyle{mystyle}{
    commentstyle=\color{brown},
    keywordstyle=\color{magenta},
    numberstyle=\tiny\color{codegray},
    stringstyle=\color{codepurple},
    basicstyle=\ttfamily\footnotesize,
    breakatwhitespace=false,         
    breaklines=true,                 
    captionpos=b,                    
    keepspaces=true,                 
    numbers=none,                    
    numbersep=5pt,                  
    showspaces=false,                
    showstringspaces=false,
    showtabs=false,                  
    tabsize=2,
    escapeinside={<@}{@>}
}

\newcommand{\tool}{\textsc{Optimo}\xspace}
\newcommand{\coffe}{\textsc{Coffe}\xspace}
\newcommand{\effibench}{\textsc{Effibench}\xspace}

\newcommand{\tabincell}[2]{\begin{tabular}{@{}#1@{}}#2\end{tabular}}
\newcommand{\eat}[1]{\if 0 #1 \fi}

\newcommand{\g}{\cellcolor{mygray}}

\newfont{\mycrnotice}{ptmr8t at 7pt}
\newfont{\myconfname}{ptmri8t at 7pt}

\newcommand{\answer}[2]{
\begin{tcolorbox}[breakable,width=\linewidth,boxrule=0pt,top=1pt, bottom=1pt, left=1pt,right=1pt, colback=gray!20,colframe=gray!20]
\textbf{Answer to RQ#1:} #2
\end{tcolorbox}
}

\begin{document}

\title{Multi-level Code Optimization via Mixture of Prompts}


\author{Yun Peng}
\orcid{0000-0003-1936-5598}
\affiliation{%
  \institution{Fudan University}
  \city{Shanghai}
  \country{China}
}
\email{yunpeng4@sigsoft.org}

\author{Jun Wan}
\orcid{0009-0006-3294-688X}
\affiliation{%
  \institution{Zhejiang University}
  \city{Hangzhou}
  \country{China}
}
\email{22451014@zju.edu.cn}

\author{Jiakun Liu}
\orcid{0000-0002-7273-6709}
\affiliation{%
  \institution{Harbin Institute of Technology}
  \city{Harbin}
  \country{China}
}
\email{jiakunliu@hit.edu.cn}

\author{Shuzheng Gao}
\orcid{0000-0002-8102-480X}
\affiliation{%
  \institution{Chinese University of Hong Kong}
  \city{Hong Kong}
  \country{Hong Kong}
}
\email{szgao23@cse.cuhk.edu.hk}

\author{David Lo}
\orcid{0000-0002-4367-7201}
\affiliation{%
  \institution{Singapore Management University}
  \city{Singapore}
  \country{Singapore}
}
\email{davidlo@smu.edu.sg}

\author{Xiaoxue Ren}
\orcid{0000-0002-5526-1617}
\authornote{Corresponding author.}
\affiliation{%
  \institution{Zhejiang University}
  \city{Hangzhou}
  \country{China}
}
\affiliation{%
  \institution{Hangzhou High-Tech Zone (Binjiang) Institute of Blockchain and Data Security}
  \city{Hangzhou}
  \country{China}
}
\email{xxren@zju.edu.cn}


\begin{abstract}
    Runtime efficiency is a critical factor that impacts both software quality and user satisfaction. There are many approaches proposed for code optimization to improve runtime efficiency. Traditional code optimization methods operate on intermediate representations (IRs) during compilation for static languages. They are effective but struggle to handle dynamic languages that do not require compilation. Recently, large language models (LLMs) have been leveraged to directly optimize source code in dynamic languages. However, these methods fail to identify suitable optimization targets and usually conduct incomprehensive single-level optimization.

    To address these challenges, we propose \tool, a multi-level LLM-based code optimization approach built on a novel Mixture-of-Prompts (MoP) architecture. In the MoP architecture, \tool identifies time-critical code structures as performance bottlenecks via differential profiling. These structures are then routed to some optimization strategies, akin to expert models in MoE, each tailored to optimize specific code patterns. Unlike traditional approaches that focus only on statement-level optimizations, \tool operates at four levels of abstraction, ranging from coarse-grained algorithmic improvements to fine-grained optimizations in API usage. We evaluate \tool on two code efficiency benchmarks, \coffe and \effibench. Our results demonstrate that \tool achieves an up to 57.48\% opt\%, i.e., the percentage of optimized programs that are correct and at least 10\% faster than the original programs, and an up to 3.97x speedup when optimizing human-written code, and it consistently outperforms the best baseline by up to 96.51\% in terms of opt\%. Furthermore, \tool achieves an up to 42.42\% opt\% and an up to 13.51x speedup when optimizing LLM-generated code.
\end{abstract}




\maketitle

\section{Introduction}\label{sec:intro}

In the ISO/IEC 25010 quality model~\cite{iso}, runtime efficiency is an important software quality attribute beyond functionality. Even when software is functionally correct, poor runtime efficiency can lead to CWE issues~\cite{cwe} and performance bugs~\cite{jovic2011catch,nistor13discovering}, which may reduce user satisfaction and even cause financial loss~\cite{garg22deepdev}. However, time inefficiencies are pervasive in real-world software, and avoiding them completely during development is difficult in practice~\cite{jin2012understanding,zhao2022large,belias2025performance}. 

Code optimization aims to automatically improve the runtime efficiency of software. Traditionally, it is integrated into the compilation process of static languages. For example, the C/C++ compiler GCC provides many optimization options~\cite{gccopt} that apply different optimization techniques~\cite{wolf1991loop,wolf1991data,cytron1991efficiently,wegman1991constant,ball1996efficient} on intermediate representations (IRs) during compilation. With the rapid growth of artificial intelligence, dynamic languages such as Python have become increasingly popular~\cite{octoverse}. Without the need for compilation, developers can quickly build functional software using these languages. However, software written in dynamic languages is often less efficient than that written in static languages, partly because it does not benefit from compiler-based optimization. For example, a recursive Python function takes more than 20 seconds to compute the Fibonacci sequence of length 40, whereas a similar C implementation takes less than one second.\footnote{Evaluated on Ubuntu 20.04 with two Intel Xeon@2.20GHZ CPUs.} Unfortunately, many traditional code optimization techniques are not applicable to dynamic languages because they rely on intermediate states generated during compilation. This limitation has motivated researchers to explore optimization techniques that work directly on source code.

Recent advances in large language models (LLMs) have shown strong capabilities in source code understanding~\cite{lichain, zhaounveiling, beger2025coconut, chen2025reasoning} and generation~\cite{wang22no,zhang23repocoder,chen24teaching,jiang23selfevolve}, leading to a growing line of work on LLM-based code optimization. For example, EffiLearner~\cite{huang2024effilearner} extracts execution traces and uses them to guide LLMs in optimizing code. RAPGEN~\cite{garg2025rapgen} reformulates code optimization as an automated program repair task and fixes inefficient code in source code. SBLLM~\cite{gao2024sbllm} combines LLMs with evolutionary search to identify suitable optimization strategies. Although these approaches have advanced the state of the art, they still face two important challenges.

\begin{figure}
    \centering
    \includegraphics[width=0.95\linewidth]{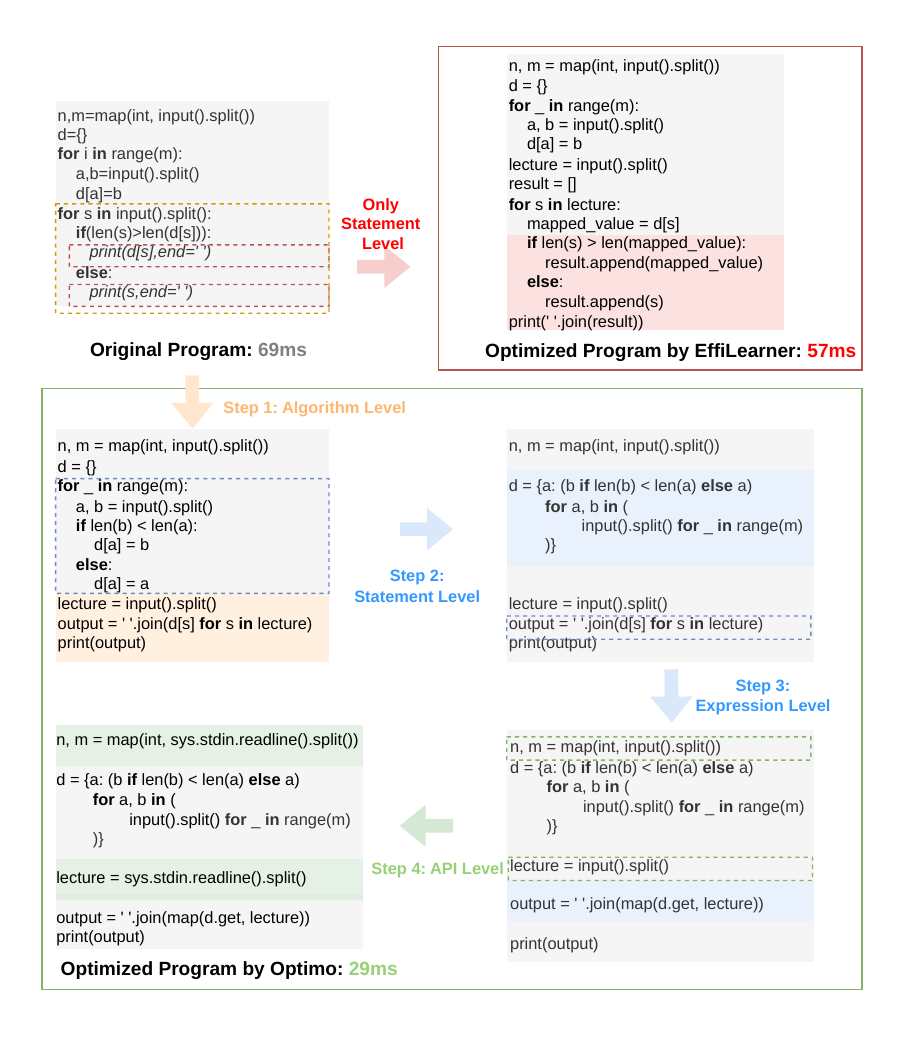}
    \caption{A motivating example from \coffe. The code in the dashed box is the optimization target for the next level, while the highlighted code is the optimized code from the identified optimization target at the last level.}
    \label{fig:mot}
\end{figure}

\textbf{Challenge 1: Failure to identify suitable optimization targets.}
Early studies~\cite{peng2024perfcodegen,shypula2024learning} optimize the entire program without first locating performance bottlenecks. More recent work, such as EffiLearner~\cite{huang2024effilearner}, relies solely on LLMs to analyze execution traces and identify lines or statements with high execution time as optimization targets. However, such lines or statements are not always appropriate targets because some operations do not meaningfully vary with input size and offer limited opportunities for optimization. For example, I/O operations, such as printing a number, often take longer than other operations, but they usually cannot be further optimized at the source code level without changing the underlying system calls. As a result, simply selecting expensive statements may mislead the optimization process.

\textbf{Challenge 2: Incomprehensive single-level optimization.} 
Recent approaches~\cite{huang2024effilearner,garg2025rapgen} mainly focus on statement-level or line-level optimization. They detect and optimize inefficiencies in several statements or lines. While this can improve efficiency to some extent, such single-level optimization is inherently limited because runtime efficiency depends on multiple factors, including algorithm design, statement and expression structure, and API usage. For example, improving a recursive implementation often requires changes to the entire program rather than small edits to a few lines of code. Without a comprehensive view of bottlenecks, optimization at a single level is often insufficient to achieve substantial performance gains.

To address these challenges, we propose \tool, a multi-level LLM-based code optimization approach built on a novel Mixture-of-Prompts (MoP) architecture. Inspired by the success of Mixture-of-Experts (MoE) architectures~\cite{shazeer2017outrageously,lepikhin2020gshard,fedus2022switch,deepseekv3} in modern LLMs, \tool replaces expert models with optimization strategy prompts that encode domain knowledge in the MoP architecture for code optimization. This eliminates the need for high-quality training data in the MoE architecture. Under the MoP architecture, \tool first generates differential test case pairs with small and large inputs, and then profiles the program to identify time-critical code structures as optimization targets. Specifically, time-critical code structures are statements or expressions whose execution time is higher than average and changes significantly across differential test cases. Such structures are suitable targets for optimization because they are more likely to have scalable and efficient alternatives. \tool then routes these targets to some optimization strategies. Optimization strategies are mined from existing slow-fast code pairs and capture rich optimization knowledge for different optimization targets. \tool applies the routed optimization strategies to the optimization targets for targeted optimization and fuses optimizations for multiple targets in the program to build an optimized program. Unlike prior approaches that only optimize at the statement level, \tool performs optimization at four levels sequentially: algorithm, statement, expression, and API. This ensures that \tool can cover both coarse-grained and fine-grained opportunities for comprehensive improvement.

We evaluate \tool on \coffe and \effibench against six LLM-based code optimization approaches. The results show that
\tool achieves an up to 57.48\% opt\%, i.e., the percentage of optimized programs that are correct and at least 10\% faster than the original programs, and an up to 3.97x speedup for human-written code optimization. It consistently outperforms the best baseline by up to 96.51\% in terms of opt\%. \tool is also highly effective on LLM-generated code optimization, delivering at least 3x higher opt\% than the baselines on \effibench. More importantly, 26.82-54.60\% of the LLM-generated code optimized by \tool is even more efficient than human-written code. Our ablation study further confirms the effectiveness of both the MoP architecture and the multi-level optimization design.

\textbf{Contributions.} We summarize our contributions as follows.
\begin{itemize}[leftmargin=*]
    \item To the best of our knowledge, \tool is the first multi-level LLM-based approach for code optimization.
    \item We propose a novel Mixture-of-Prompts architecture for code optimization. It uses differential profiling to identify time-critical code structures as optimization targets and routes them to mined optimization strategies for targeted optimization.
    \item Extensive experiments demonstrate the superior performance of \tool on both human-written and LLM-generated code optimization, and verify the effectiveness of the MoP architecture and the multi-level optimization design.
\end{itemize}
\section{Motivation}\label{sec:mot}

\begin{figure*}[t]
    \centering
    \includegraphics[width=0.9\linewidth]{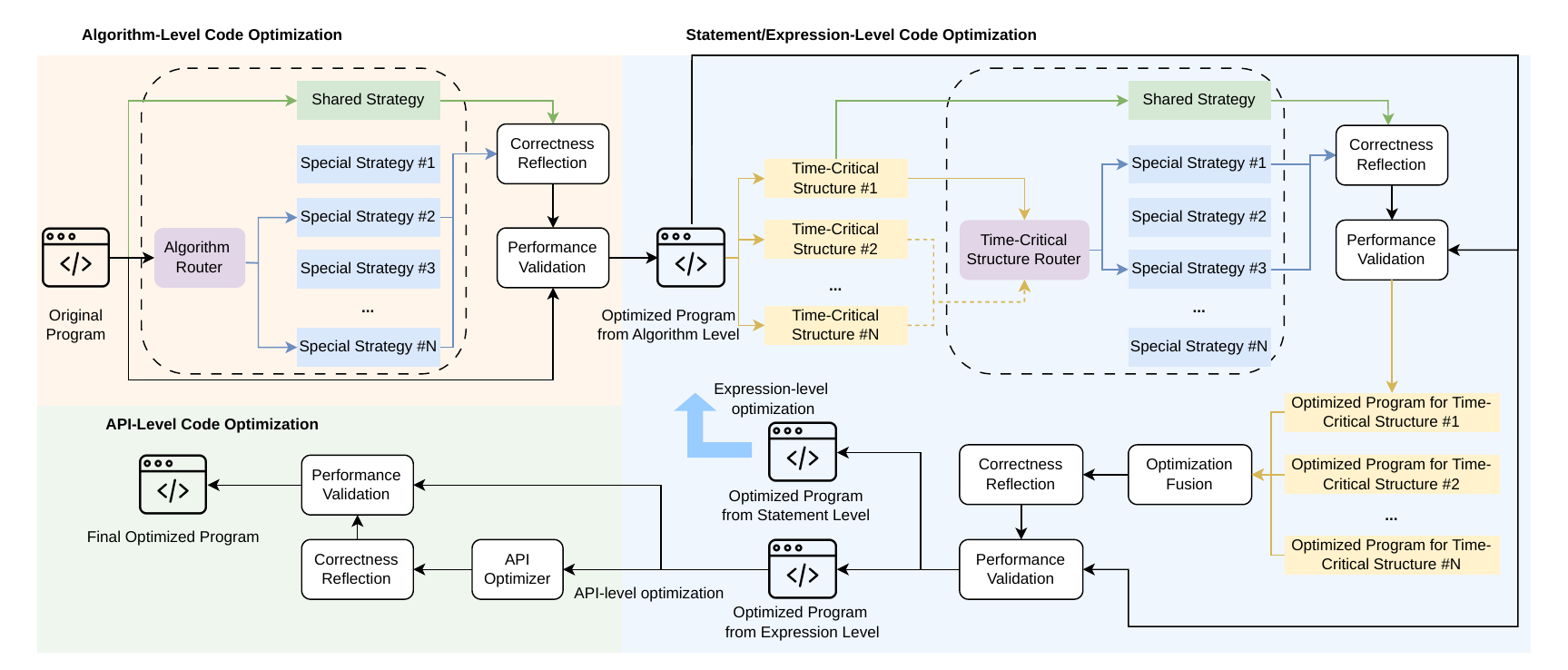}
    \caption{The overview of \tool. It optimizes a program at four levels sequentially: algorithm, statement, expression, and API. At each level except API, the MoP architecture highlighted in the dashed box performs targeted optimization.}
    \label{fig:overview}
\end{figure*}

We provide an example in Fig.~\ref{fig:mot} to better illustrate the challenges of current approaches and how \tool addresses them. 

\textbf{Baseline.} EffiLearner is a state-of-the-art LLM-based approach for code optimization. Given the original program, it runs it on some test cases and identifies two \texttt{print} statements with the highest execution time as performance bottlenecks. It then optimizes the two \texttt{print} statements by moving them out of the loop, as highlighted in red. The final optimized program takes 57ms. Even though it is faster than the original program, the speedup is only 1.21x. There are two reasons for the limited improvement. Firstly, while \texttt{print} statements take the most time, the two \texttt{for} loops are the root causes for the inefficiency. Incorrectly identified bottlenecks mislead LLMs during the optimization process (Challenge 1). Secondly, EffiLearner focuses only on a few expensive statements, ignoring algorithm-level optimizations to reduce the overall time complexity (Challenge 2).

\textbf{Our Approach.} Our approach \tool implements four optimization levels to obtain a comprehensive improvement. It first analyzes the time complexity of the original program at the algorithm level, and reduces it from O(m + n) to O(m), as highlighted in orange. It then applies differential profiling and identifies that the execution time of the \texttt{for} loop varies the most. \tool replaces it with a list comprehension, as highlighted in blue. At the expression level, \tool identifies that the list comprehension to calculate the value of \texttt{output} can be further replaced with a more efficient \texttt{map} function. Finally, \tool replaces the expensive \texttt{input} calls with \texttt{readline} calls at the API level. Benefiting from the four levels of optimization, the program optimized by \tool runs in only 29ms, achieving a 2.38x speedup, significantly faster than the program optimized by EffiLearner. Furthermore, \tool correctly identifies \texttt{for} loops for optimization based on differential profiling. The MoP architecture, with its rich domain knowledge, also ensures that \tool can implement more efficient modifications than EffiLearner.
\section{Methodology}\label{sec:meth}


\subsection{Mixture-of-Prompts (MoP) Architecture}

Fig.~\ref{fig:overview} illustrates the workflow of \tool. Given a program, \tool performs code optimization at four levels sequentially, i.e., algorithm, statement, expression, and API. We determine the four levels based on the node levels in abstract syntax trees (ASTs), from the root node representing the entire program (algorithm level) to the leaf node representing API calls (API level). This design reduces the risk that changes made at earlier levels are completely overwritten by later optimizations. \tool addresses the limitation of single-level optimization by covering both high-level functionality and low-level code structures. At each level except the API level, \tool employs a novel Mixture-of-Prompts (MoP) architecture for targeted optimization. Note that \tool does not implement the MoP architecture for the API level, since inefficient APIs can be directly replaced with efficient alternatives via an efficient API map.

Mixture-of-Experts (MoE) architectures~\cite{shazeer2017outrageously,lepikhin2020gshard,fedus2022switch,deepseekv3} in modern LLMs have demonstrated superior performance across diverse tasks. Despite their effectiveness, they cannot be directly applied to code optimization due to a lack of high-quality slow-fast code data for training. Inspired by the MoE architecture, \tool introduces a MoP architecture for code optimization. At each level, \tool identifies optimization targets and routes them to the appropriate optimization strategies. Specifically, \tool identifies algorithms and time-critical code structures through differential profiling, based on the insight that code structures whose cost grows significantly with input scale are more likely to admit more efficient variants. This design helps \tool focus on targets that offer meaningful optimization opportunities, addressing the challenge of previous approaches in identifying inappropriate optimization targets.

Instead of using multiple expert models that require training, \tool employs multiple optimization strategies that can be obtained in a small set of slow-fast code pairs, including \textit{special optimization strategies} and \textit{shared optimization strategies}. Special optimization strategies are specialized prompts tailored to optimize a specific algorithm or code structure under different conditions. \tool mines them from slow-fast code pairs in existing datasets. These strategies capture domain knowledge about how human developers optimize specific code elements in practice. Shared optimization strategies are general optimization prompts we manually defined for each level. They build the basic performance of \tool by leveraging LLMs' reasoning capabilities, and handle cases not covered by special optimization strategies. For the edits generated by each optimization strategy, \tool validates both correctness and runtime efficiency, and then merges the validated edits into the original program to produce the optimized program at each level.

\subsection{Time-Critical Code Structure Identification}\label{sec:identification}

We define time-critical code structures as statements or expressions whose execution time (1) exceeds the average execution time of a statement in the program and (2) changes the most across different inputs. Such code structures not only contribute substantially to the total runtime but are also more likely to benefit from optimization under different input scales. Note that this process does not apply at the algorithm level, as it pertains only to a few code elements in the program. Instead, we use time complexity to characterize the program's overall performance.

\textbf{Differential Profiling.} Given a program, \tool first generates small and large test cases with inputs of different scales. \tool adopts the stress test case generation approach STGen~\cite{peng2025coffe} to generate large test cases. \tool directly uses public test cases as small test cases if available. Otherwise, it uses a modified version of STGen to generate small test cases. All test cases are validated against the program to ensure correctness. \tool executes the program and applies LineProfiler~\cite{lineprofiler} to collect line-by-line execution information on all test cases, i.e., the time ratio, total time, execution count, and time per execution for each line.

\textbf{Time-Critical Statement and Expression Identification.} 
Based on execution information collected through differential profiling, \tool first identifies lines whose total time ratio exceeds the program average in both small and large test cases, indicating that they are critical to time efficiency under any circumstance. It then compares the time ratios of these lines across small and large test cases, and selects those whose ratios increase on large test cases. We regard these lines as time-critical because they are both expensive and sensitive to input scale, offering more optimization opportunities. 

Next, \tool classifies time-critical lines into potential time-critical statements or expressions. If the increased runtime is primarily due to a higher execution count, the line is treated as a time-critical statement. If it is mainly caused by increased time per execution, the line is treated as a time-critical expression. These candidates are then converted into ASTs by selecting the smallest subtree that contains them. In this way, both time-critical statements and expressions are represented as subtrees. For time-critical statements, \tool keeps only statement-level AST nodes. For time-critical expressions, it keeps only expression-level AST nodes.

\subsection{Optimization Strategy Mining}

Apart from the optimization targets, optimization strategies are also important components in the MoP architecture of \tool. Among them, shared optimization strategies are manually defined to handle general cases, while special optimization strategies encode domain knowledge from developers for code optimization. Unlike the expert models in the MoE architecture, which require larger training datasets, \tool mines special optimization strategies from a few slow-fast code pairs.

\subsubsection{Data Preparation}
Programming competition datasets are a useful source for mining optimization strategies because they often contain multiple solutions to the same problem with varying runtime performance. In this work, we use a Python Codeforces submission dataset~\cite{codeforces} from HuggingFace. Its training split contains 621,000 Python solutions for a wide range of Codeforces problems. We clean this dataset by keeping only correct solutions that pass all test cases and grouping them by problem. For each problem, we then select solutions whose execution time differs significantly from that of other solutions. Following prior work~\cite{shypula2024learning,gao2024sbllm}, we treat a runtime difference greater than 10\% as significant. Problems with fewer than two valid solutions are removed. After data cleaning, 98\% of solutions are removed, leaving only 13,319 valid code solutions across 2,005 problems, with an average of 6.64 per problem.

To identify time-critical code structures in these solutions, we treat the original test cases in the dataset as small test cases and use STGen~\cite{peng2025coffe} to generate large test cases. We then run the solutions on the large test cases and rank them by execution time. Finally, for each problem, we pair the fastest solution with the remaining slower solutions to construct slow-fast code pairs.

\subsubsection{Optimization Strategy Mining}

Given the slow-fast code pairs, \tool performs three phases to mine special optimization strategies at the algorithm, statement, and expression levels. 

\begin{figure}[t]
    \centering
    \includegraphics[width=1.0\linewidth]{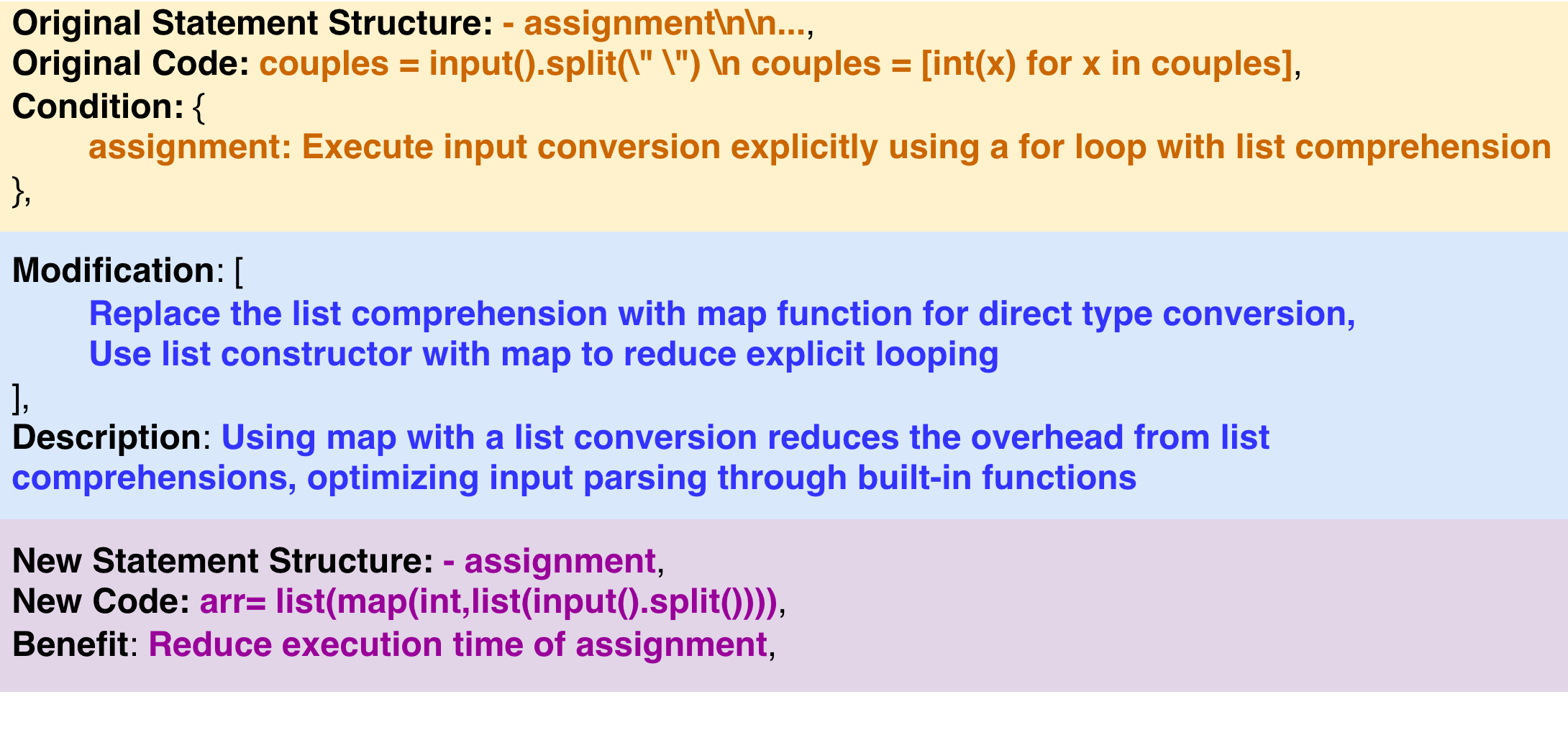}
    \caption{An initial optimization strategy extracted from a time-critical statement in a fast-slow code pair.}
    \label{fig:opt}
\end{figure}

\textbf{Phase I: Optimization Strategy Extraction.} For each slow-fast code pair, \tool identifies the time complexity of the slow code using LLMs at the algorithm level, and time-critical code structures using differential profiling at the statement and expression levels. These are treated as optimization targets at their corresponding levels. Given the identified targets in the slow code, \tool then asks LLMs to analyze how these targets are transformed into more efficient versions in the fast code, and extracts an initial optimization strategy that describes the transformation.

Fig.~\ref{fig:opt} shows an example of an initial optimization strategy at the expression level. It contains three parts:
\begin{itemize}[leftmargin=*]
    \item \textbf{Routing Pattern} (highlighted in yellow): an optimization target for syntactic routing, and a condition for semantic routing, where the optimization target is an algorithm or code structure to be optimized, and the condition specifies when the strategy applies if there are multiple strategies for the same optimization target.
    \item \textbf{Optimization Method} (highlighted in blue): a modification and a description, where the modification describes how to transform the target into a more efficient version, and the description provides a high-level summary to help better understanding.
    \item \textbf{Optimization Result} (highlighted in purple): an optimized target and a benefit, where the optimized target is the algorithm or code structure after applying the optimization method, and the benefit states how runtime efficiency is improved.
\end{itemize}

From the 13,319 cleaned code solutions, \tool extracts 2,942, 11,172, and 13,885 initial optimization strategies at the algorithm, statement, and expression levels, respectively. The initial optimization strategies document how developers optimize specific programs under specific circumstances.

\textbf{Phase II: Optimization Method Summarization.} Due to the similar methodologies developers use to optimize programs, there may be multiple initial optimization strategies with the same routing pattern, and each individual strategy may be too specific to generalize across different scenarios. To improve generalizability, \tool summarizes the optimization methods of optimization strategies with the same routing pattern. Specifically, it groups all initial strategies by the routing pattern, and then uses LLMs to summarize one optimization method in each group by proposing broader modification steps. If more than 20 groups remain for an optimization target, \tool further merges strategies with similar conditions to reduce the number of groups to 20. This avoids overloading the MoP architecture with too many optimization strategies for a single optimization target.

\textbf{Phase III: Optimization Target Abstraction.} Due to the limited volume of slow-fast code pairs, initial optimization strategies are not enough to cover all code patterns. Furthermore, code structures follow a long-tail distribution, and some complex structures are covered by only a single initial optimization strategy, which can hardly generalize across different scenarios. To address this issue, \tool abstracts complex code structures and enables them to share optimization strategies from simpler ones. 

As described in Sec.~\ref{sec:identification}, \tool represents time-critical code structures as subtrees in AST. Subtrees with greater depth have more nodes and are thus harder to match. Inspired by prior work~\cite{peng24domain}, \tool simplifies subtrees by gradually removing leaf nodes, thereby discarding some details while preserving the overall structure. For example, we can remove the node of an assignment statement in the \texttt{for} loop statement to produce a simpler subtree without changing the loop structure. Therefore, for all time-critical code structures associated with only one initial optimization strategy, \tool reduces their depth by one to create simplified structures. It then shares the optimization strategies of the simplified structures. This abstraction process can be repeated multiple times until simplified structures are found.

The second and third phases may iterate several times until all initial optimization strategies are transformed into general optimization strategies that cover multiple scenarios. These strategies are used as special optimization strategies in the MoP architecture of \tool. We finally obtained 191 special optimization strategies for 48 time complexities at the algorithm level, 851 at the statement level for 272 time-critical statements, and 805 at the expression level for 272 time-critical expressions.

In addition, we manually design a shared optimization strategy for each level. Specifically, we follow the chain-of-thought prompt design from prior work~\cite{shypula2024learning} and ask LLMs to optimize the target solely based on their reasoning capabilities. Shared optimization strategies ensure that every optimization target is handled appropriately, even when no special optimization strategy applies.

\subsubsection{Efficient API Mining}
Compared with algorithms and code structures, APIs are often too fine-grained and diverse to be comprehensively covered by existing coding datasets. Therefore, \tool identifies efficient APIs based on API documentation and LLMs. Specifically, it first collects the APIs used in the coding dataset and retrieves relevant APIs from the corresponding documentation to form an initial API set. It then groups APIs by functionality and asks LLMs to enumerate as many alternative APIs as possible within each group. We refer to APIs as functionally equivalent if they share the same inputs and outputs, as measured by test cases. Based on the expanded API set, \tool generates test cases using LLMs to evaluate runtime efficiency and ranks the APIs in each group accordingly. \tool discards groups in which the runtime difference between the fastest and slowest APIs is not significant (i.e., less than 10\%). Finally, \tool constructs an efficient API map covering 685 groups.

\subsection{Multi-level Code Optimization}
With optimization targets as the router and optimization strategies for targeted optimization, \tool is now able to optimize programs based on the MoP architecture. As shown in Fig.~\ref{fig:overview}, \tool starts with coarse-grained algorithm-level optimization and ends with fine-grained API-level optimization. This order avoids later optimizations from overwriting earlier ones.

\subsubsection{Algorithm-level Code Optimization} 

As highlighted in orange in Fig.~\ref{fig:overview}, given a program, \tool analyzes its time complexity and applies several special optimization strategies based on the time complexity, along with a shared optimization strategy for general optimization. \tool then applies the optimization methods in the selected optimization strategies to the program based on LLMs and generates multiple optimized programs as candidates. However, \tool does not directly adopt these candidates as final optimized programs, since previous work has shown that code optimization can easily break functional correctness~\cite{peng2024perfcodegen}. To mitigate this problem, \tool introduces a correctness reflection process to verify candidate programs. The verified candidates, along with the original program, are then ranked by execution time, and the most efficient program is selected for further optimization at the next level.

\textbf{Correctness Reflection.} In this process, \tool checks the functional correctness of candidate programs using the small test cases generated in Sec.~\ref{sec:identification}. For incorrect candidates, \tool applies a reflection process that uses failed test cases and corresponding error messages as feedback to repair them. Only repaired or previously verified correct candidates are retained, while the remaining incorrect ones are discarded. This is a self-reflection process: \tool relies only on LLM-generated test cases rather than external feedback. Correctness reflection provides two benefits. First, it performs early filtering, removing incorrect candidates before they enter later optimization levels. Second, it provides correctness feedback, helping optimization strategies preserve the original program's functionality. Note that \tool does not assume that small test cases are sufficient to identify all incorrect programs, and the reflection process is not always successful. Therefore, this process is neither complete nor sound.

\textbf{Performance Validation.} After correctness reflection, \tool ranks the correct candidate programs by execution time on the large test cases. The original program is also included in this ranking to guard against performance regression. A candidate program is selected for the next level only if it is faster than the original program. Otherwise, the original program is retained, and all candidate programs are discarded. In this way, performance validation serves as a gatekeeper, ensuring that the program passed to the next level is at least as efficient as the input program.

\subsubsection{Statement-level and Expression-level Code Optimization} Given the optimized program from the algorithm level, \tool identifies time-critical statements and time-critical expressions for statement-level and expression-level optimization, respectively. The optimization procedure is the same at both levels, as highlighted in blue in Fig.~\ref{fig:overview}, but the corresponding optimization strategies differ. Unlike algorithm-level optimization, where each program has only one time complexity, a program may contain multiple time-critical code structures. \tool therefore optimizes them in parallel and merges the edits afterward. To achieve this, \tool creates an optimization pipeline for each time-critical code structure and executes these pipelines in parallel. In each pipeline, \tool complements the optimization process at the algorithm level by adding an adaptive routing process to improve the coverage.

\textbf{Adaptive Routing.} \tool uses an adaptive routing method to map rarely appearing time-critical code structures to optimization strategies. Given a code structure, \tool first matches it to optimization strategies with exactly the same optimization target. If no strategy is matched, \tool abstracts the time-critical code structure into a more general one by reducing its depth. This abstraction continues until one or more special optimization strategies are found. Among the selected strategies, \tool compares their conditions against the semantics of the time-critical code structure and removes inappropriate ones. Adaptive routing improves coverage by allowing complex code structures to fall back to more optimization strategies with simpler optimization targets.

Within each pipeline, \tool generates candidate programs and applies the same correctness reflection and performance validation process used at the algorithm level. Each pipeline then outputs a ranked list of candidates for the corresponding time-critical code structure. To integrate optimizations across multiple code structures, \tool asks LLMs to select suitable candidates and merge them into a single program. The integrated programs are also subjected to correctness reflection and performance validation before being passed to the next level.

\subsubsection{API-level Code Optimization} At this level, \tool identifies inefficient APIs used in the program and replaces them with more efficient alternatives based on the efficient API map. Because API-level changes usually have limited impact on program functionality, \tool does not create parallel optimization pipelines at this level. Instead, it iteratively replaces all inefficient APIs and applies the same correctness reflection and performance validation processes. The optimized program is regarded as the final output of \tool.
\section{Experiment Setup}\label{sec:setup}

\subsection{Research Questions}
To evaluate the effectiveness of \tool, we study the following research questions:
\begin{itemize}[leftmargin=*]
    \item \textbf{RQ1:} To what extent can the optimization strategies mined by \tool cover time-critical code structures in human-written and LLM-generated code?
    \item \textbf{RQ2:} How effective is \tool compared with existing approaches in optimizing human-written code?
    \item \textbf{RQ3:} How effective is \tool compared with existing approaches in optimizing LLM-generated code?
    \item \textbf{RQ4:} What are the impacts of different components in \tool?
    \item \textbf{RQ5:} What are the costs of \tool?
\end{itemize}

\subsection{Baselines}
We select the following approaches in LLM-based code optimization as our baselines:
\begin{itemize}[leftmargin=*]
    \item \textbf{In-Context Learning (ICL) Prompt}: We use the same in-context learning prompting setting as PIE~\cite{shypula2024learning}, where several slow-fast code pairs are used as demonstrations.
    \item \textbf{Chain-of-Thought (CoT) Prompt}: We use the same chain-of-thought prompting strategy as PIE~\cite{shypula2024learning}.
    \item \textbf{PIE}~\cite{shypula2024learning}: PIE proposes a fine-tuning framework for code optimization. We use the fine-tuned model released by the authors as a baseline.
    \item \textbf{SBLLM}~\cite{gao2024sbllm}: It is a search-based LLM framework that iteratively refines candidate solutions and discovers better strategies for code optimization.
    \item \textbf{EffiLearner}~\cite{huang2024effilearner}: It is a self-optimization framework that leverages execution overhead profiles to improve runtime efficiency with LLMs.
    \item \textbf{RAPGEN}~\cite{garg2025rapgen}: It is a retrieval-augmented generation approach for fixing code inefficiency issues.
\end{itemize}

\begin{table}[t]
    \centering
    \caption{The coverage of special optimization strategies in \tool for time-critical statements and expressions in human-written and LLM-generated code. ``Func'' indicates the \coffe-Function and ``File'' indicates \coffe-File. ``Method'' indicates the method used to select optimization strategies for time-critical code statements and expressions.}
    \scalebox{0.75}{
    \begin{tabular}{cccccccc}
    \toprule
      \multirow{2}*{\textbf{Method}}  & \multirow{2}*{\textbf{Source}}   &  \multicolumn{3}{c}{\textbf{Statement}} & \multicolumn{3}{c}{\textbf{Expression}} \\
      \cmidrule{3-8}
      & &  \textbf{Func} &\textbf{ File} & \textbf{Effibench} &  \textbf{Func} & \textbf{File} &\textbf{ Effibench} \\
      \midrule
      \multirow{2}*{\tabincell{c}{\textbf{Exact }\\\textbf{Match}}}& Human & 73.68\%  & 69.31\% & 60.17\% & 90.91\% & 91.23\% & 87.34\% \\
      & GPT-4o & 68.36\% & 80.26\% & 72.62\% & 89.26\% & 98.18\% & 85.13\% \\
      \midrule
      \multirow{2}*{\tabincell{c}{\textbf{Adaptive} \\ \textbf{Routing}}} & Human & 89.47\% &	97.88\% &	98.79\% &	96.36\% &	99.68\% &	99.37\% \\
      & GPT-4o & 87.57\% &	98.68\% &	99.51\% &	94.21\% &	100.00\% &	98.83\% \\
      \bottomrule
    \end{tabular}}
    \label{tab:rq1_hit}
\end{table}

\subsection{Benchmark}
We evaluate \tool on \coffe~\cite{peng2025coffe} and \effibench~\cite{effibench}, two benchmarks for evaluating the runtime efficiency for code generation and optimization approaches. \coffe contains 398 function-level problems and 294 file-level problems, while \effibench contains 707 file-level problems. Note that \effibench originally included 1,000 problems, but we retain only the 707 problems whose ground-truth solutions strictly pass all provided test cases. We also remove 64 duplicate problems in \coffe that appear in the Codeforces datasets or \effibench, to avoid data contamination and potential bias in evaluation.

\subsection{Metrics}
We adopt the metrics from previous work~\cite{shypula2024learning,gao2024sbllm} on evaluation:
\begin{itemize}[leftmargin=*]
    \item \textbf{Correct\%:} The percentage of optimized programs that pass all test cases.
    \item \textbf{Opt\%:} The percentage of optimized programs $o$ that are correct and at least 10\% faster than the original programs $s$, i.e.,  $\frac{t_s - t_o}{t_o} > 10\%$, where $t_s$ and $t_o$ denote the execution time of the original and optimized programs, respectively. We use 10\% as the threshold to maintain consistency with prior work~\cite{shypula2024learning,gao2024sbllm}.
    \item \textbf{Speedup:} The ratio $\frac{t_s}{t_o}$, where $t_s$ is the execution time of the original program and $t_o$ is the execution time of the optimized program, when the optimized program is faster. Otherwise, the speedup is set to 1, indicating that optimization fails to improve runtime efficiency. In addition to the per-instance speedup, we also compute the overall speedup as $\frac{\sum_1^N t_s}{\sum_1^N t_o}$, which measures the aggregate speedup achieved by an approach over the entire benchmark. Compared with per-instance speedup, overall speedup provides a more stable benchmark-level evaluation.
\end{itemize}

\begin{table*}[t]
\centering
\caption{The evaluation results of \tool and baselines on two benchmarks. ``Speedup'' indicates the overall speedup of the entire benchmark. Numbers in \colorbox{mygray}{gray} indicate the best performance.}
\scalebox{0.85}{
\begin{tabular}{lccccccccc}
\toprule
& \multicolumn{3}{c}{\textbf{COFFE-Function}} & \multicolumn{3}{c}{\textbf{COFFE-File}}  & \multicolumn{3}{c}{\textbf{Effibench}}     
\\
\midrule
\textbf{Approach}    & \multicolumn{1}{c}{\textbf{Correct\%}} & \multicolumn{1}{c}{\textbf{Opt\%}}   & \textbf{Speedup} & \multicolumn{1}{c}{\textbf{Correct\%}} & \multicolumn{1}{c}{\textbf{Opt\%}}  & \textbf{Speedup} & \multicolumn{1}{c}{\textbf{Correct\%}} & \multicolumn{1}{c}{\textbf{Opt\%}}  & \textbf{Speedup} \\ \midrule
\textbf{ICL}         & 70.60\%                                & 33.67\%                            & 2.67                                 & 77.21\%                                & 22.45\%                            & 1.11                                 & 45.54\%                                & 5.37\%                             & 1.01                                 \\
\textbf{CoT}         & 64.57\%                                & 35.43\%                            & 2.61                                 & 66.33\%                                & 21.09\%                            & 1.12                                 & 55.02\%                                & 7.21\%                             & 1.04                           \\
\textbf{EffiLearner} & 84.92\%                          & 44.72\%                      & 2.73                                 & 87.41\%                          & 29.25\%                      & 1.13                           & 51.91\%                                & 11.32\%                      & 1.04                           \\
\textbf{PIE}         & 82.41\%                                & 36.18\%                            & 2.17                                 & 80.27\%                                & 15.31\%                            & 1.08                                 & 77.09\%                          & 2.40\%                             & 1.01                                 \\
\textbf{RAPGEN}      & 78.64\%                                & 37.94\%                            & 3.30                           & 79.93\%                                & 22.45\%                            & 1.12                                 & 63.93\%                                & 5.80\%                             & 1.04                           \\
\textbf{SBLLM}       & 65.83\%                                & 30.15\%                            & 2.09                                 & 62.24\%                                & 14.29\%                            & 1.02                                 & 48.66\%                                & 3.82\%                             & 1.02                                 \\
\midrule
\textbf{Ours}        & \g 98.74\%                       & \g 45.48\%                   & \g 3.97                        & \g 98.30\%                       & \g 57.48\%                   & \g 1.33                        & \g 85.15\%                       & \g 16.12\%                   & \g 1.10       
\\ \bottomrule
\end{tabular}}
\label{tab:rq2_main}
\end{table*}

\subsection{Implementation}

For all baselines and \tool, we use GPT-4o~\cite{gpt4o} as the base model unless otherwise specified, to ensure a fair comparison. We adopt the evaluation framework from \coffe~\cite{peng2025coffe}. Specifically, each code solution is executed 12 times in independent Docker containers, and the average of the middle 10 runs is used as the final execution time. For all runtime measurements, we use wall-clock time, which captures the total time a program takes to execute, including CPU computation, I/O operations, and other system activities. All experiments are conducted on a Linux machine running Ubuntu 20.04, equipped with two Intel Xeon@2.20GHz CPUs and 1 TB RAM.

\section{Experiment Results}\label{sec:eval}

\subsection{RQ1: Coverage of Time-Critical Code Structures}

\tool optimizes time-critical code structures mainly by optimization strategies, so it is important to evaluate the coverage of these strategies. We first identify time-critical code structures in code written by both developers and GPT-4o across the two benchmarks, and then measure how many of these structures are covered by the optimization targets of optimization strategies in \tool. The results are shown in Table~\ref{tab:rq1_hit}.

From Table~\ref{tab:rq1_hit}, we observe that the optimization strategies in \tool can strictly cover 60.17-80.26\% of time-critical statements and 85.13-98.18\% of time-critical expressions under exact matching on the two benchmarks. When adaptive routing is enabled, coverage further increases to 87.57-99.51\% and 94.21-100\% for time-critical statements and expressions, respectively. These results suggest that time-critical code structures are highly repetitive across programs, and that the special optimization strategies mined from 2,005 problems are sufficient to cover most of them with adaptive routing. In other words, \tool does not require an extremely large dataset to mine effective and broadly applicable optimization strategies. In addition, \tool achieves consistently high coverage across human-written and LLM-generated code, demonstrating its ability to handle both types of programs effectively.

\answer{1}{With adaptive routing, \tool covers 87.57-100\% of time-critical code structures in both human-written and LLM-generated code.}

\subsection{RQ2: Effectiveness on Human-written Code}

In RQ2, we evaluate how effectively \tool optimizes human-written code. Specifically, we compare the correctness and runtime efficiency achieved by \tool and the baselines on human-written ground-truth solutions from the two benchmarks.

\begin{figure}[t]
    \centering
    \subfigure[Per-instance speedup distribution on \coffe-Func.]{
    \begin{minipage}[t]{0.53\linewidth}
    \centering
    \includegraphics[width = 1.0\textwidth]{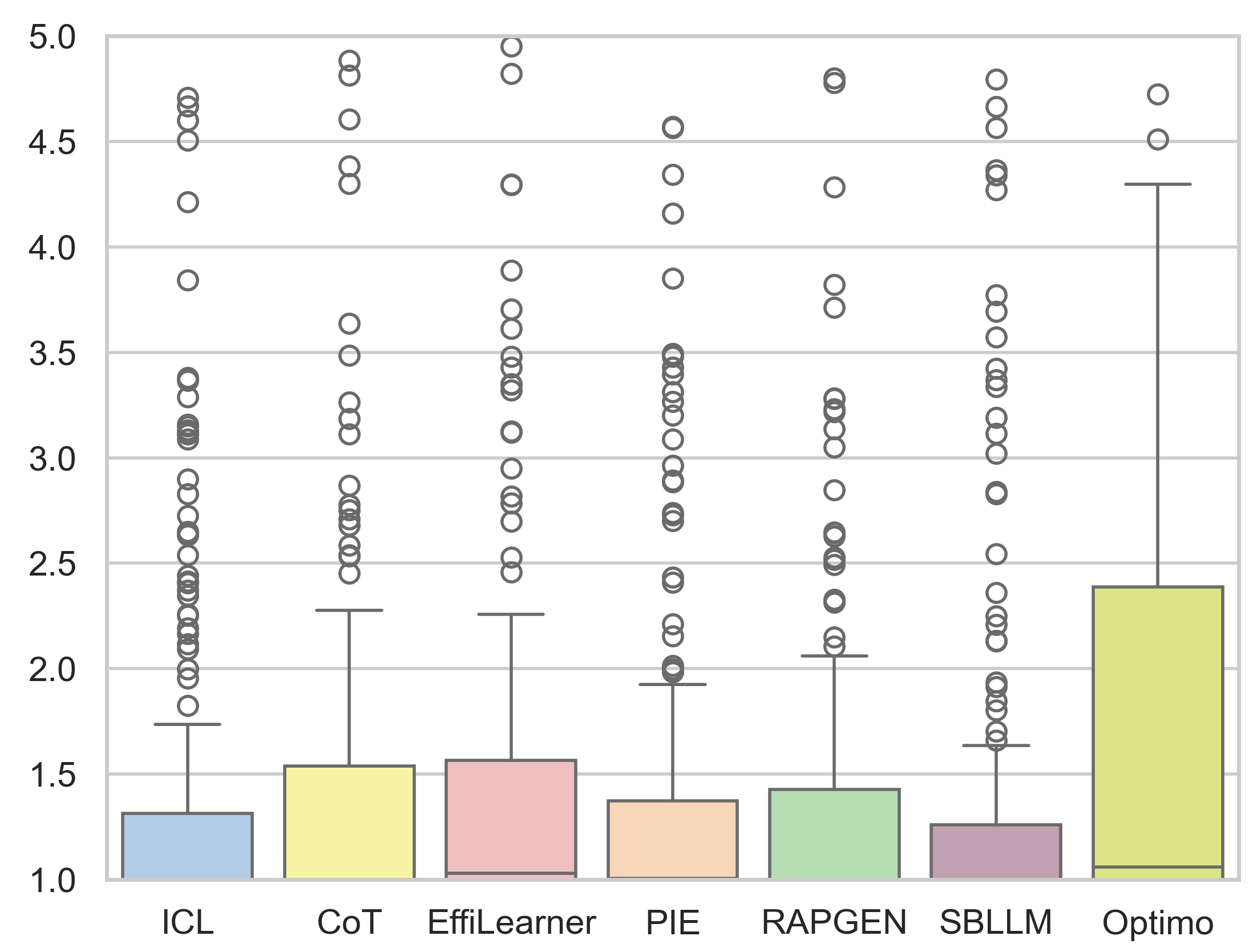}
    \label{fig:func_speedup}
    \end{minipage}
    }
    \subfigure[Venn diagram on \coffe-Func.]{
    \begin{minipage}[t]{0.42\linewidth}
    \centering
    \includegraphics[width = 1.0\textwidth]{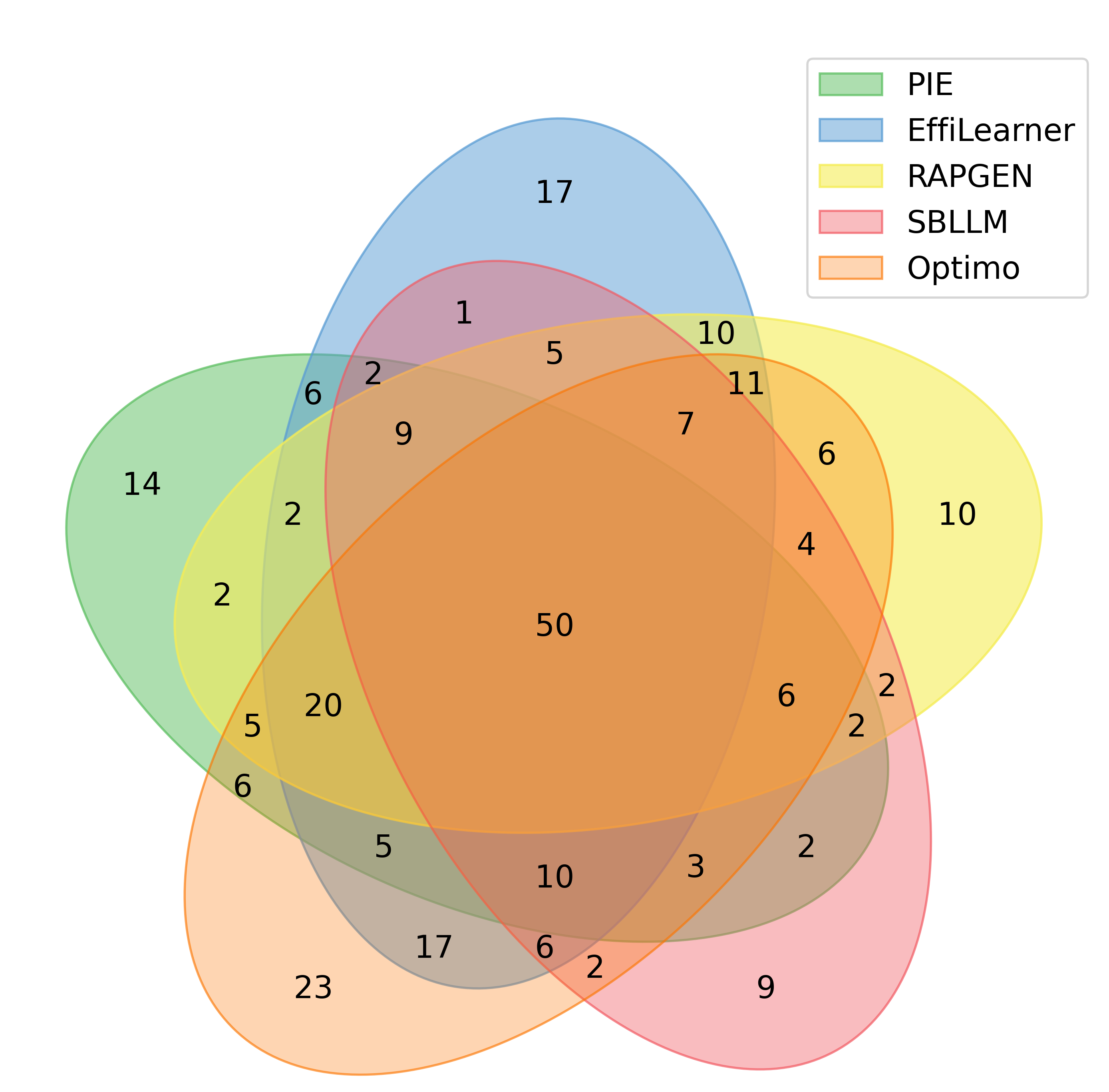}
    \label{fig:func_venn}
    \end{minipage}
    }
    \subfigure[Per-instance speedup distribution on \coffe-File.]{
    \begin{minipage}[t]{0.53\linewidth}
    \centering
    \includegraphics[width = 1.0\textwidth]{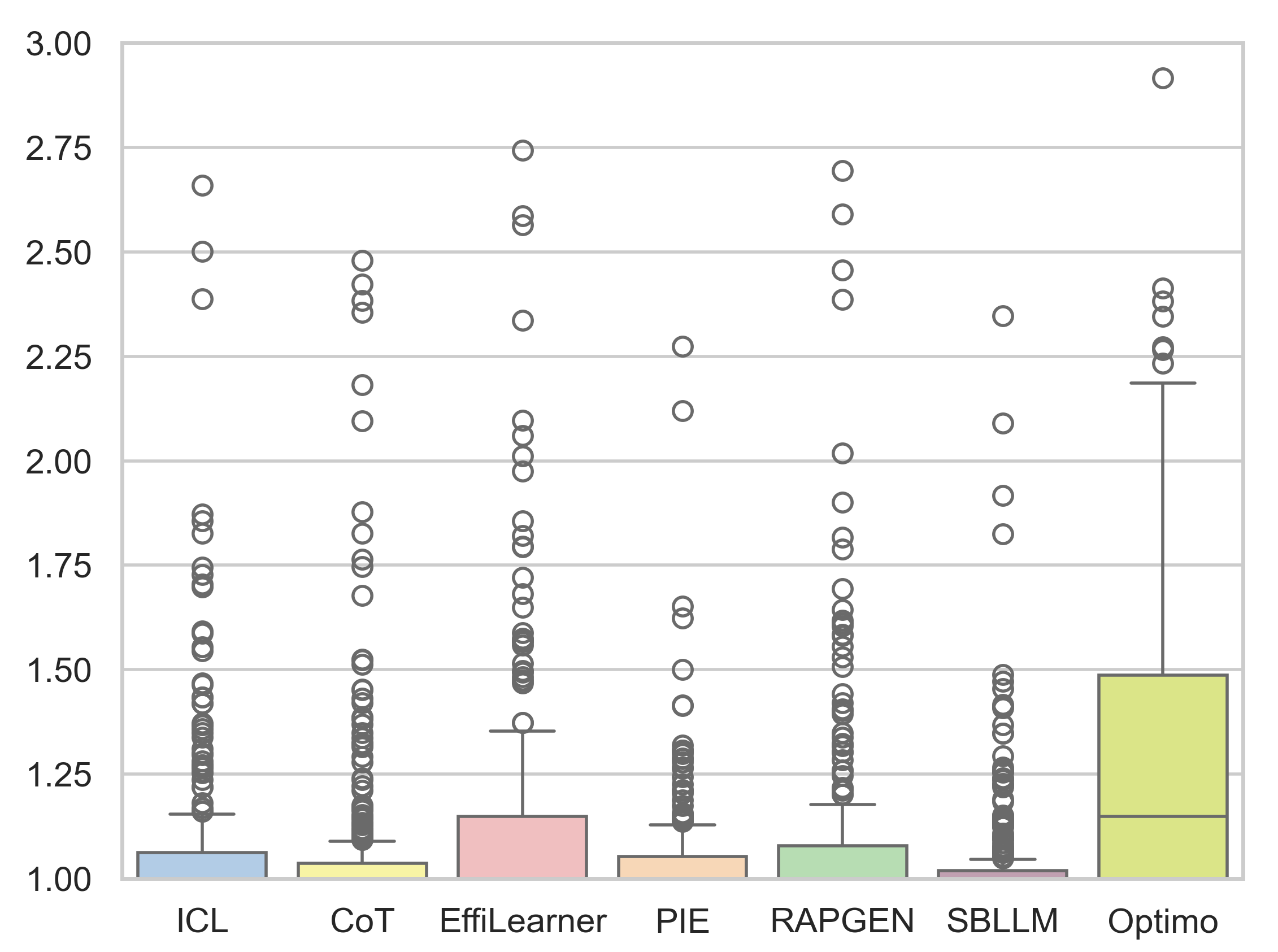}
    \label{fig:file_speedup}
    \end{minipage}
    }
    \subfigure[Venn diagram on \coffe-File.]{
    \begin{minipage}[t]{0.42\linewidth}
    \centering
    \includegraphics[width = 1.0\textwidth]{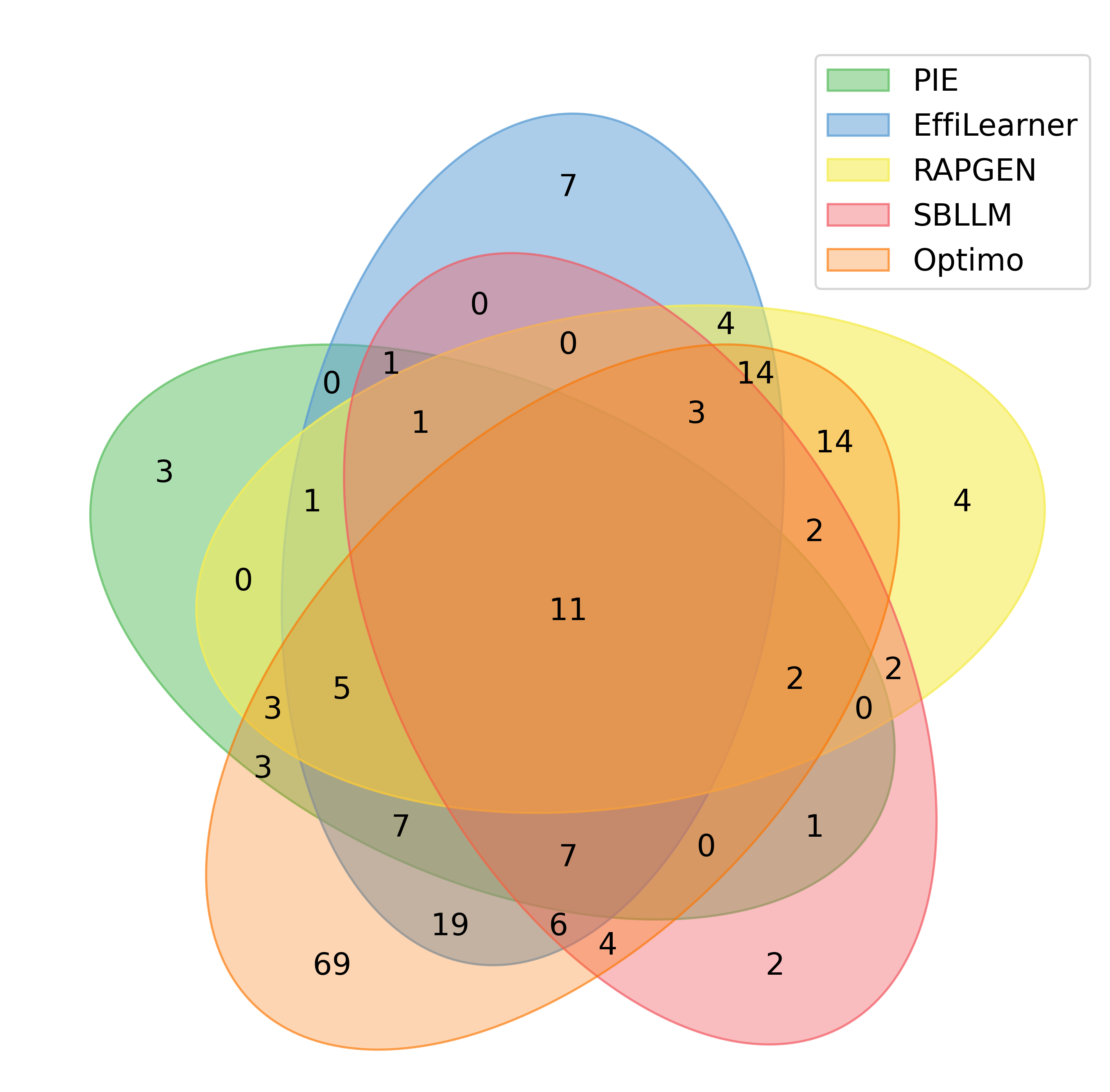}
    \label{fig:file_venn}
    \end{minipage}
    }
    \subfigure[Per-instance speedup distribution on \effibench.]{
    \begin{minipage}[t]{0.53\linewidth}
    \centering
    \includegraphics[width = 1.0\textwidth]{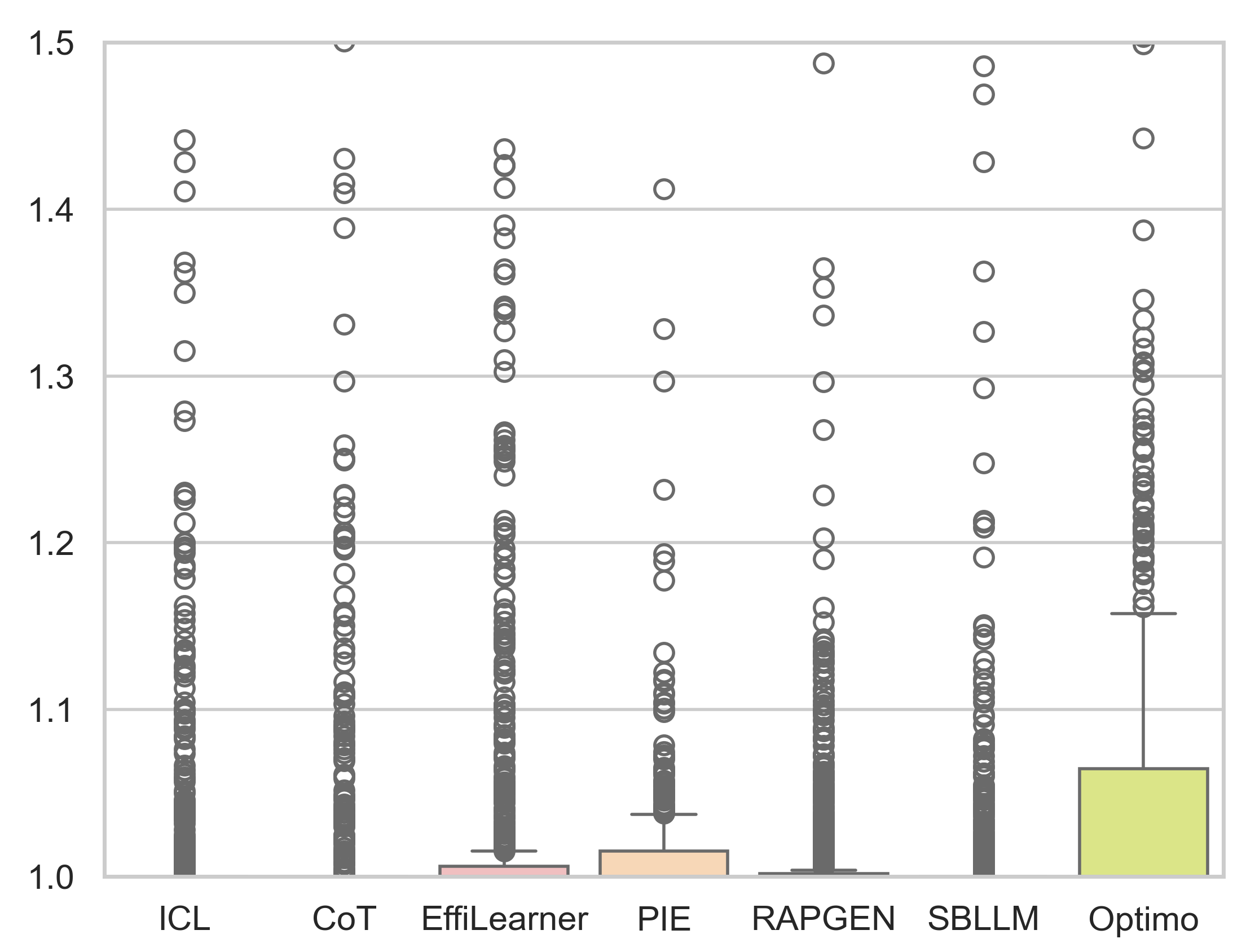}
    \label{fig:effibench_speedup}
    \end{minipage}
    }
    \subfigure[Venn diagram on \effibench.]{
    \begin{minipage}[t]{0.42\linewidth}
    \centering
    \includegraphics[width = 1.0\textwidth]{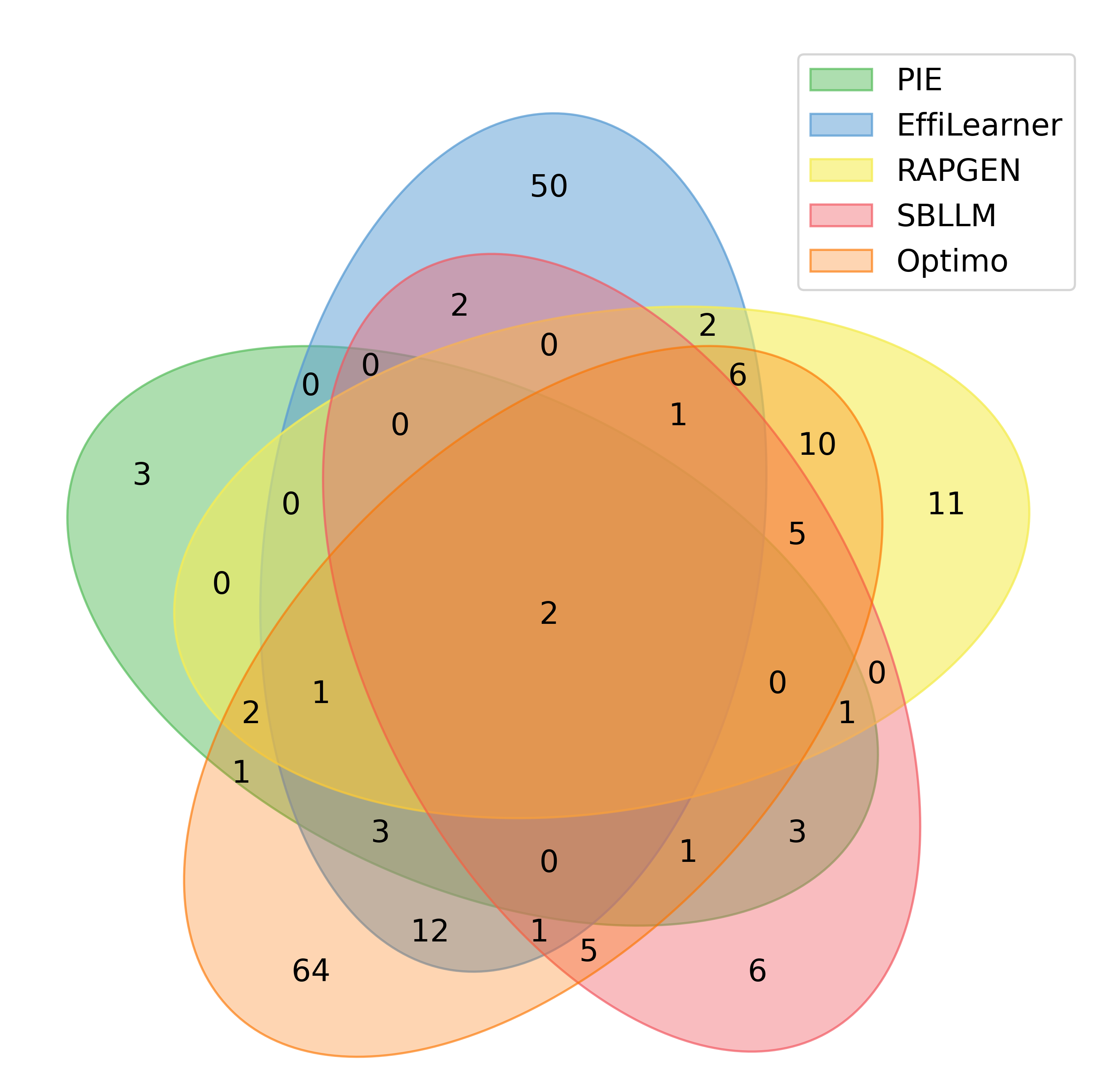}
    \label{fig:effibench_venn}
    \end{minipage}
    }
    \caption{The per-instance speedup distribution and Venn Diagrams of optimized problems of \tool and baselines on \coffe and \effibench.}
    \label{fig:rq2_fig}
\end{figure}

\textbf{Overall Performance.} Table~\ref{tab:rq2_main} presents the results of all baselines and \tool. To calculate the overall speedup in the table, we use the execution time of the original program for incorrectly or slower optimized programs, since the original program is always available even when the optimized version is incorrect or slower.

From Table~\ref{tab:rq2_main}, we observe that \tool consistently achieves the best correct\%, outperforming the strongest baseline by 16.27\%, 12.46\%, and 10.46\% on \coffe-Function, \coffe-File, and \effibench, respectively. We attribute this improvement to the correctness reflection process in \tool, which reuses the test cases generated during differential profiling to preserve program correctness after optimization. We also admit that the correctness reflection cannot guarantee a 100\% correct\% due to the limited number of test cases.

\tool also delivers substantial gains in opt\% and speedup over existing approaches. Although all optimization strategies are mined from file-level datasets, \tool still achieves the best performance on \coffe-Function, reaching 45.48\% opt\% and an overall speedup of 3.97. This result demonstrates the generalizability of the optimization strategies used in \tool. On file-level benchmarks, \tool outperforms all existing approaches by a large margin, achieving improvements of at least 96.51\% and 42.40\% in opt\% on \coffe-File and \effibench, respectively. We further note that these gains cannot be explained solely by improvements in correct\%. For example, EffiLearner achieves a similar correct\% of 87.41\% on \coffe-File, but its opt\% is only about half that of \tool. Similarly, PIE achieves a correct\% of 77.09\% on \effibench, yet its opt\% is only 2.40\%. These findings further highlight the effectiveness of the MoP architecture and the multi-level optimization design in \tool.

\textbf{Per-instance Speedup Distribution.} Beyond overall speedup, we also analyze the distribution of per-instance speedup on the two benchmarks. The corresponding boxplots are shown in Fig.~\ref{fig:func_speedup}, Fig.~\ref{fig:file_speedup}, and Fig.~\ref{fig:effibench_speedup}. Following prior work~\cite{gao2024sbllm}, we set the speedup of incorrectly optimized instances to 1. From Fig.~\ref{fig:func_speedup} and Fig.~\ref{fig:file_speedup}, we observe that \tool produces more optimized programs with larger speedups. In particular, about 25\% of the optimized programs produced by \tool achieve a speedup of 2.5x on \coffe-Function and 1.5x on \coffe-File. In contrast, the maximum speedup achieved by the baselines, excluding outliers, is even lower than 2.5x and 1.5x on the corresponding benchmarks. Fig.~\ref{fig:effibench_speedup} further shows that \tool can still improve a substantial number of instances on \effibench, even though most baselines make little or no improvement. To assess whether the observed differences are statistically significant, we compare \tool with each baseline using Wilcoxon signed-rank tests~\cite{wilcoxon1992individual} on paired per-instance speedups. The results show that \tool significantly outperforms every baseline on both benchmarks (p < 0.05) and has positive effect sizes in every comparison. The per-instance speedup distributions confirm that the effectiveness of \tool stems from improvements across many instances rather than relying on a few extreme cases.

\textbf{Uniquely Optimized Instances.} We further analyze the instances that can be significantly improved by the baselines and \tool, and present the corresponding Venn diagrams in Fig.~\ref{fig:func_venn}, Fig.~\ref{fig:file_venn}, and Fig.~\ref{fig:effibench_venn}. For clarity, we show only the four strongest baselines in these figures. Across all three benchmarks, \tool uniquely optimizes 23, 69, and 64 instances on \coffe-Function, \coffe-File, and \effibench, respectively, whereas the best baseline uniquely optimizes only 17, 7, and 50 instances. This result suggests that the superior performance of \tool cannot be obtained by simply combining existing approaches. By identifying more suitable optimization targets and applying targeted multi-level optimization under the MoP architecture, \tool can optimize instances that remain difficult for prior code optimization methods.

\answer{2}{\tool consistently improves both correct\% and opt\% on the two benchmarks, outperforming the strongest baselines by up to 16.27\% and 96.51\%, respectively. It also achieves the highest overall speedup and better per-instance speedup distribution by uniquely optimizing the largest number of instances.}

\subsection{RQ3: Effectiveness on LLM-generated Code}

\begin{table*}[t]
\centering
\caption{The evaluation results of \tool and baselines on correct solutions generated by GPT-4o on two benchmarks. All columns and labels are the same as Table~\ref{tab:rq2_main}.}
\scalebox{0.85}{
\begin{tabular}{lccccccccc}
\toprule
& \multicolumn{3}{c}{\textbf{COFFE-Function}} & \multicolumn{3}{c}{\textbf{COFFE-File}}  & \multicolumn{3}{c}{\textbf{Effibench}}     
\\
\midrule
\textbf{Approach}    & \multicolumn{1}{c}{\textbf{Correct\%}} & \multicolumn{1}{c}{\textbf{Opt\%}}   & \textbf{Speedup} & \multicolumn{1}{c}{\textbf{Correct\%}} & \multicolumn{1}{c}{\textbf{Opt\%}}  & \textbf{Speedup} & \multicolumn{1}{c}{\textbf{Correct\%}} & \multicolumn{1}{c}{\textbf{Opt\%}}  & \textbf{Speedup} \\ \midrule
\textbf{ICL}         & 59.93\%                                & 19.52\%                            & 1.76                                 & 66.29\%                                & 9.09\%                             & 1.01                                 & 76.34\%                                & 2.98\%                             & 1.01                                 \\
\textbf{CoT}         & 75.68\%                                & 28.77\%                            & 1.73                                 & 61.74\%                                & 12.12\%                            & 1.04                                 & 65.61\%                                & 3.58\%                             & 1.01                                 \\
\textbf{EffiLearner} & 90.07\%                                & 34.59\%                            & 7.15                                 & 81.06\%                                & 27.65\%                            & 1.06                                 & 60.04\%                                & 11.73\%                            & 1.15                                 \\
\textbf{PIE}         & 89.38\%                                & 24.66\%                            & 1.58                                 & 68.18\%                                & 7.95\%                             & 1.02                                 & 89.86\%                                & 1.99\%                             & 1.03                                 \\
\textbf{RAPGEN}      & 78.42\%                                & 23.63\%                            & 1.13                                 & 67.42\%                                & 14.77\%                            & 1.04                                 & 53.28\%                                & 8.15\%                             & 1.00                                 \\
\textbf{SBLLM}       & 69.52\%                                & 19.86\%                            & 1.60                                 & 62.12\%                                & 12.12\%                            & 1.01                                 & 77.14\%                                & 3.58\%                             & 1.00                                 \\
\midrule
\textbf{Ours}        & \g 96.67\%                                & \g 41.78\%                            & \g 13.51                                & \g 98.48\%                                & \g 42.42\%                            & \g 1.08                                 & \g 97.42\%                                & \g 35.79\%                            & \g 6.41                                                            
   \\ \bottomrule
\end{tabular}}
\vspace{-10pt}
\label{tab:rq3_main}
\end{table*} 

\begin{figure*}[t]
    \subfigure[\coffe-Function]{
    \begin{minipage}[t]{0.32\linewidth}
        \centering
        \includegraphics[width = 1.05\textwidth]{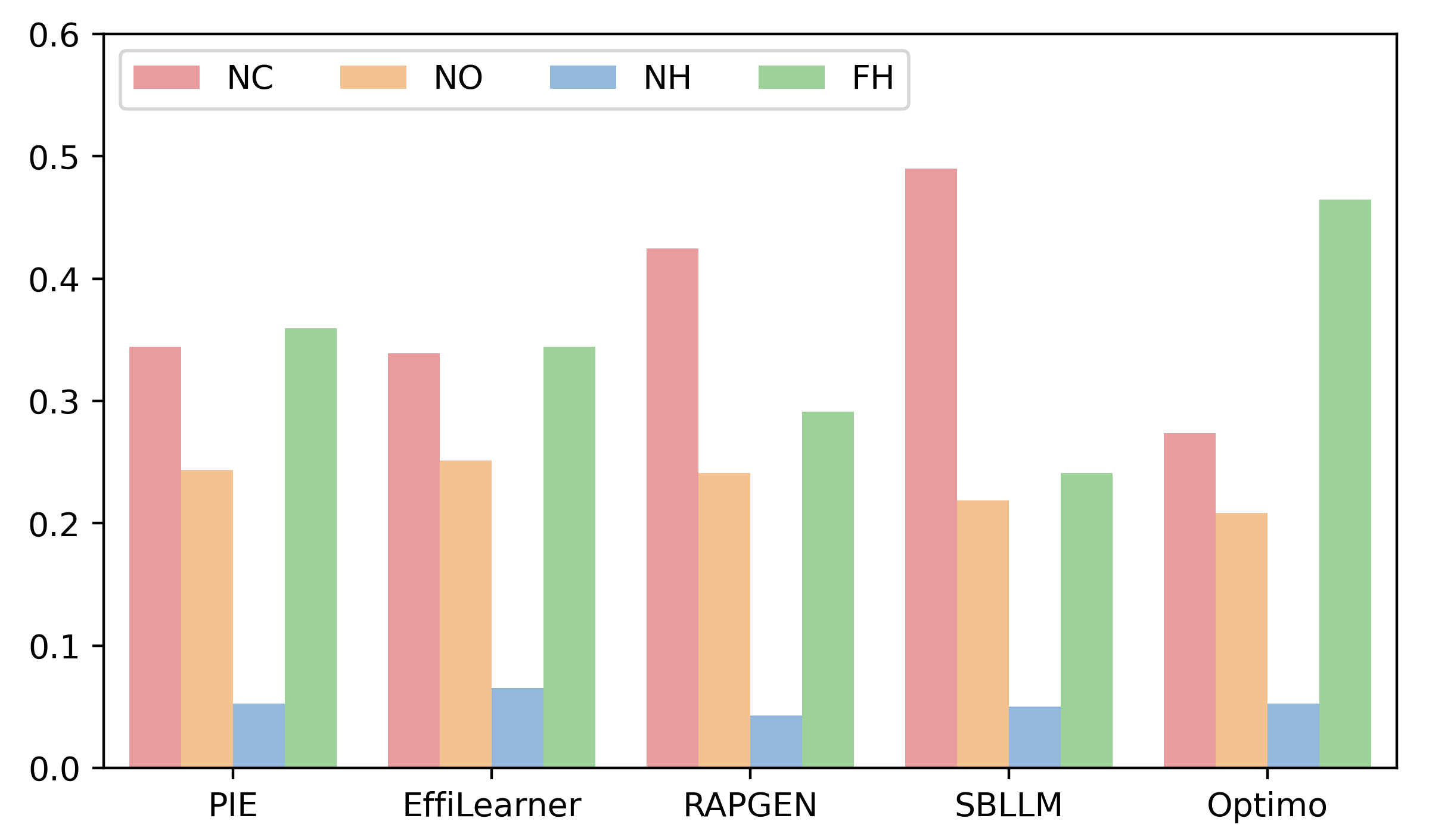}
        \label{fig:function_barplot}
    \end{minipage}
    }
    \subfigure[\coffe-File]{
    \begin{minipage}[t]{0.32\linewidth}
        \centering
        \includegraphics[width = 1.05\textwidth]{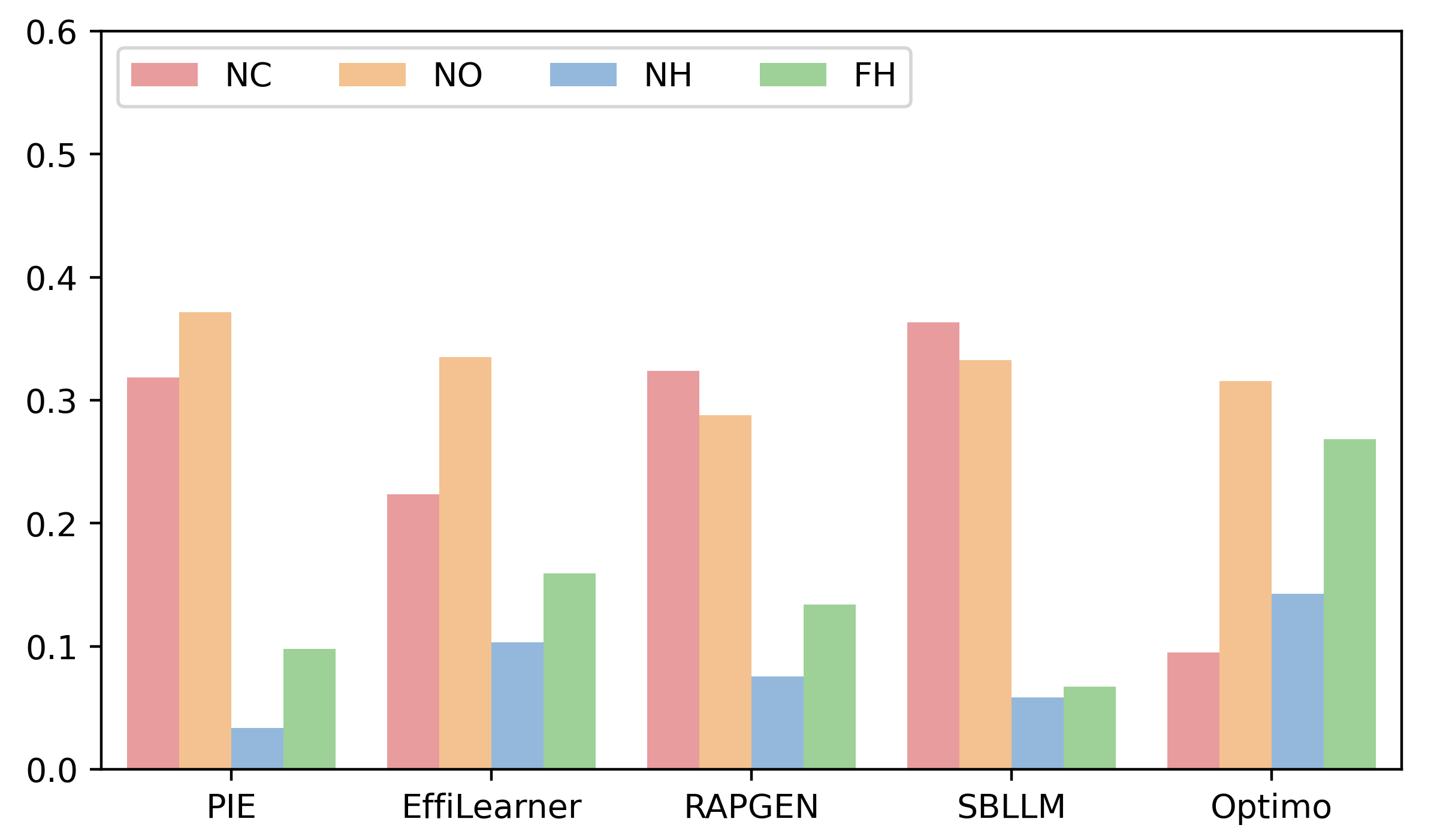}
        \label{fig:file_barplot}
    \end{minipage}
    }
    \subfigure[\effibench]{
    \begin{minipage}[t]{0.32\linewidth}
        \centering
        \includegraphics[width = 1.06\textwidth]{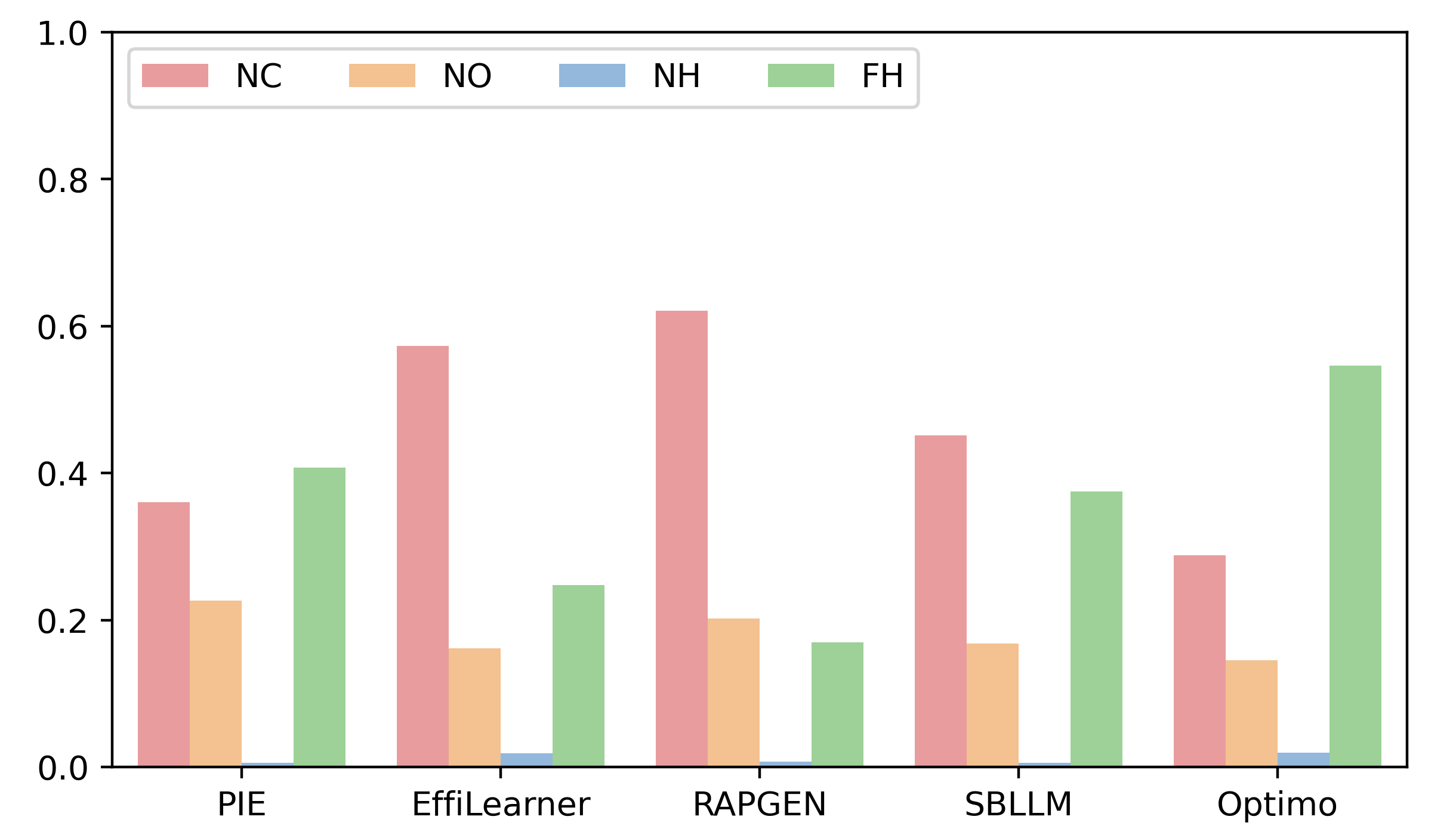}
        \label{fig:effibench_barplot}
    \end{minipage}
    }
    \caption{The distribution of optimized LLM-generated solutions on two benchmarks. ``NC'', ``NO'', ``NH'', ``FH'' indicate the final optimized solution is ``not correct'', ``correct but not optimized'', ``optimized but not faster than human solutions'', and ``faster than human solutions'', respectively.} 
    \label{fig:rq3_human}
\end{figure*}

To answer RQ3, instead of directly using ground truth solutions in the two benchmarks, we use GPT-4o to generate solutions at zero temperature for optimization. After validating the generated solutions on the test cases, we obtain 292, 264, and 503 correct solutions as the original programs for \coffe-Function, \coffe-File, and \effibench, respectively. We employ \tool and baselines to optimize these programs.

\textbf{Overall Performance.} We report the results in Table~\ref{tab:rq3_main}. Firstly, we observe that simple prompting approaches such as ICL and CoT achieve much lower opt\% on LLM-generated code than on human-written code. For example, ICL achieves an opt\% of only 19.52\% on LLM-generated code in \coffe-Function in Table~\ref{tab:rq3_main}, compared with 33.67\% on human-written code in Table~\ref{tab:rq2_main}. In contrast, more advanced approaches, such as EffiLearner, achieve relatively stable performance across both settings. This suggests that simple prompting approaches alone are often insufficient to further optimize code generated by the same LLM, as they rely solely on LLMs' internal capabilities for code optimization. Advanced approaches remain effective since they usually incorporate domain knowledge and external feedback.

By leveraging rich domain knowledge from optimization strategies, \tool achieves the best performance in most settings. For example, it obtains an overall speedup of 13.51 and 6.41 on \coffe-Function and \effibench, respectively, substantially outperforming all baselines. In addition, \tool achieves an opt\% of 35.79\% on \effibench, which is at least three times higher than that of any baseline and nearly twice the opt\% achieved by \tool on human-written code. These results demonstrate the strong ability of \tool to optimize LLM-generated code.

\textbf{Comparison with Human-written Solutions.} We further study whether a program fully generated and optimized by LLMs can be as efficient as a program written by developers. To compare the runtime efficiency of LLM-generated code and human-written code optimized by \tool and baselines, we categorize the optimized programs of \tool and the baselines into four groups, following prior work~\cite{gao2024sbllm}. ``NC'' includes code that is either incorrectly generated by GPT-4o or incorrectly optimized by the optimization approach. ``NO'' includes correctly optimized code that is not faster than the original code generated by GPT-4o. ``NH'' includes correctly optimized code that is faster than the original code generated by GPT-4o but still slower than human-written code. ``FH'' includes correctly optimized code that is faster than human-written code. Fig.~\ref{fig:rq3_human} shows the proportion of instances in each category. Due to space constraints, we present only the four strongest baselines in the figure.

From Fig.~\ref{fig:rq3_human}, we observe that \tool consistently achieves the highest FH rate of 46.48\%, 26.82\%, and 54.60\% on \coffe-Function, \coffe-File, and \effibench, respectively. This indicates that \tool produces the most programs that are more efficient than human-written ones. Benefiting from the correctness reflection process, \tool also achieves the lowest NC rate across the benchmarks.

\answer{3}{\tool is also highly effective at optimizing LLM-generated code. On \effibench, it achieves an opt\% of 35.79\%, which is at least three times higher than all baselines, and 26.82-54.60\% of the optimized LLM-generated solutions are even more efficient than human-written code.}

\subsection{RQ4: Ablation Study}

To answer RQ4, we examine the effects of two major components of \tool: the multi-level optimization design and the MoP architecture. For the multi-level optimization design, we remove each optimization level in turn and evaluate the resulting variants. For the MoP architecture, we remove either the shared optimization strategies or the special optimization strategies to assess their contributions. We also remove the correctness reflection process to study its impact. The ablation results are reported in Table~\ref{tab:rq4:ablation}. Due to the computational cost, we conduct these experiments only on \coffe.

\textbf{Multi-level Optimization.} Table~\ref{tab:rq4:ablation} shows that removing any level leads to a decrease in both opt\% and speedup. This confirms the necessity of a multi-level optimization design and reinforces the limitations of existing single-level optimization approaches. Even removing the finest-grained API level causes opt\% to drop by 4.97\% and 11.24\% on \coffe-Function and \coffe-File, respectively. We further find that the largest drop occurs when removing the algorithm level on \coffe-Function (27.64\%) and the statement level on \coffe-File (23.07\%). This suggests that no single optimization level is sufficient for all programs. Instead, the full multi-level optimization design in \tool covers a broader range of optimization scenarios and helps explain its consistent superiority across benchmarks.

\textbf{Mixture-of-Prompts Architecture.} As the core of \tool, the MoP architecture also makes a substantial contribution. Removing either shared optimization strategies or special optimization strategies causes a clear decrease in opt\%, showing that both components are important. Shared optimization strategies leverage the reasoning ability of LLMs, while special optimization strategies provide domain knowledge derived from developer-written optimizations. More importantly, removing special optimization strategies leads to a much larger drop in overall speedup (e.g., from 3.97 to 2.39 on \coffe-Function) than removing shared optimization strategies (from 3.97 to 3.45 on \coffe-Function). This indicates that although LLMs have some inherent ability to optimize code, they still struggle to produce highly efficient solutions without domain knowledge from human-written examples. We also find that the correctness reflection process contributes substantially to the final opt\% because it repairs some efficient but initially incorrect optimizations, turning them into correctly optimized programs.

\begin{table}[t]
\centering
\caption{The ablation results of \tool on \coffe. All columns are the same as Table~\ref{tab:rq2_main}. ``Shared'' and ``Special'' indicate the shared and special optimization strategies.}
\scalebox{0.75}{ 
\begin{tabular}{lcccccc}
\toprule
& \multicolumn{3}{c}{\textbf{COFFE-Function}}  & \multicolumn{3}{c}{\textbf{COFFE-File}}  \\
\midrule
\textbf{Ablation}       & \multicolumn{1}{c}{\textbf{Correct\%}} & \multicolumn{1}{c}{\textbf{Opt\%}} & \multicolumn{1}{c}{\textbf{Speedup}} & \multicolumn{1}{c}{\textbf{Correct\%}} & \multicolumn{1}{c}{\textbf{Opt\%}} & \multicolumn{1}{c}{\textbf{Speedup}} \\
\midrule
\textbf{w/o Algorithm}  & 98.74\%                                & 32.91\%                            & 2.66                                 & \g 99.32\%                                & 51.36\%                            & 1.25                                 \\
\textbf{w/o Statement}  & 98.99\%                                & 39.45\%                            & 3.57                                 & 98.98\%                                & 44.22\%                            & 1.21                                 \\
\textbf{w/o Expression} & 98.74\%                                & 43.72\%                            & 3.97                                 & 98.98\%                                & 46.26\%                            & 1.25                                 \\
\textbf{w/o API}        & 98.74\%                                & 43.22\%                            & 3.86                                 & 98.30\%                                & 51.02\%                            & 1.24                                 \\ \midrule
\textbf{w/o Fix}        & 98.99\%                                & 44.22\%                            & 3.61                                 & 98.64\%                                & 48.30\%                            & 1.31                                 \\
\textbf{w/o Share}      & 97.74\%                                & 37.69\%                            & 3.45                                 & 98.30\%                                & 48.98\%                            & 1.24                                 \\
\textbf{w/o Special}    & \g 99.25\%                                & 38.69\%                            & 2.39                                 & 98.64\%                                & 47.28\%                            & 1.18                                 \\
\midrule
\textbf{\tool}           & 98.74\%                                & \g 45.48\%                            & \g 3.97                                 & 98.30\%                                & \g 57.48\%                            & \g 1.33   \\
\bottomrule
\end{tabular}}
\label{tab:rq4:ablation}
\end{table}

\answer{4}{The multi-level optimization process consistently improves the performance of \tool across different scenarios, contributing 3.87-23.07\% gains in opt\%. The MoP architecture further improves speedup by combining LLM reasoning with domain knowledge derived from developer optimizations.}

\subsection{RQ5: Cost Analysis}

To answer RQ5, we compare the inference cost of \tool and the baselines using the average numbers of input tokens, output tokens, and LLM API calls per instance. We use these model-independent metrics because monetary cost varies across model providers and deployment settings. The results on \coffe are shown in Table~\ref{tab:rq5_cost}.

\begin{table}[t]
\centering
\caption{The average inference cost of \tool and baselines on \coffe. ``Input'' and ``Output'' indicate the average numbers of input and output tokens per program, respectively. ``Calls'' indicates the average number of LLM API calls per program.}
\scalebox{0.85}{
\begin{tabular}{lcccccc}
\toprule
& \multicolumn{3}{c}{\textbf{COFFE-Function}} & \multicolumn{3}{c}{\textbf{COFFE-File}}
\\
\midrule
\textbf{Approach} & \multicolumn{1}{c}{\textbf{Input}} & \multicolumn{1}{c}{\textbf{Output}} & \textbf{Calls} & \multicolumn{1}{c}{\textbf{Input}} & \multicolumn{1}{c}{\textbf{Output}} & \textbf{Calls} \\ \midrule
\textbf{ICL}         & 2,752.50 & 314.53   & 1.00 & 1,989.60  & 408.50   & 1.00 \\
\textbf{CoT}         & 750.88   & 823.60   & 2.00    & 787.85    & 840.26   & 2.00    \\
\textbf{EffiLearner} & 755.30   & 856.07   & 2.00    & 515.67    & 125.89   & 2.00    \\
\textbf{PIE}         & 125.65 & 303.61 & 1.00 & 161.78 & 113.70 & 1.00 \\
\textbf{RAPGEN}      & 2,924.37 & 323.16   & 1.00 & 2,385.67  & 203.36   & 1.00 \\
\textbf{SBLLM}       & 839.58   & 1,171.60 & 2.83    & 601.43    & 1,408.06 & 2.85    \\
\midrule
\textbf{Ours}        & 9,521.17 & 2,076.49 & 8.46    & 20,386.64 & 3,813.09 & 16.69   \\
\bottomrule
\end{tabular}}
\label{tab:rq5_cost}
\end{table}

\textbf{Overall Cost.} Table~\ref{tab:rq5_cost} shows that \tool consumes more tokens and API calls than the baselines. This overhead mainly comes from its four-level optimization process, which generates and validates multiple candidate programs. However, the additional cost is accompanied by stronger optimization performance. For example, on \coffe-File, \tool consumes 2.71 times as many output tokens as SBLLM but achieves approximately four times its opt\% (57.48\% vs. 14.29\% in Table~\ref{tab:rq2_main}). These results demonstrate the trade-off between inference cost and optimization effectiveness on \tool.

\answer{5}{\tool incurs a higher inference cost than the baselines due to its multi-level optimization process, but achieves substantially stronger optimization performance.}
\section{Threats to Validity}\label{sec:discussion}

\textbf{LLM Selection and Data Leakage.} Due to computational resource constraints, we use GPT-4o as the base model for both \tool and all applicable baselines in our evaluation. Although \tool and most baselines are model-agnostic, their performance may vary when instantiated with other LLMs. We use the same base model for all approaches to ensure a fair comparison. In addition, there is a risk of data leakage if the benchmarks used in this paper are included in the LLM training data. To reduce this risk, we choose GPT-4o, which was released in May 2024, earlier than or close to the release dates of \coffe and \effibench. We also avoid using more recent models, which may have a higher risk of benchmark contamination.

\textbf{Generalizability to Other Programming Languages.} Although code optimization is an important task across many dynamic languages, this work focuses on Python because it is the most widely studied language in this line of research. The performance of \tool and the baselines may differ on other programming languages. To mitigate this threat, \tool is designed without components that are tightly coupled to Python-specific features. In principle, it can be extended to other languages as long as suitable datasets are available for mining optimization strategies. We leave a comprehensive evaluation of additional languages for future work.
\section{Related Work}\label{sec:literature}

\subsection{Code Optimization}
Code optimization is essential for improving runtime efficiency. Early approaches~\cite{nistor13discovering,toffola15performance,krishna20cadet,giavrimis21genetic} mainly relied on predefined rules and optimization patterns to address issues such as software misconfigurations and loop inefficiencies. With the rise of deep learning, several studies~\cite{garg22deepdev, chen22learning, chen24supersonic,du24mercury} began to learn optimization patterns from slow-fast code pairs. PIE~\cite{shypula2024learning} is a representative example, which learns performance-improving edits from slow-fast C/C++ and Python code pairs.

More recently, researchers have explored the use of LLMs for more effective code optimization~\cite{peng2024perfcodegen,ren25peace}. RAPGEN~\cite{garg2025rapgen} formulates code optimization as a program repair task and fixes inefficiency issues in zero-shot for C\# programs. EffiLearner~\cite{huang2024effilearner} leverages lightweight profiling information to iteratively guide LLMs in improving code efficiency. SBLLM~\cite{gao2024sbllm} further improves optimization performance through a search-based framework that iteratively refines the initial results generated by LLMs. Although these approaches have shown promising results, they often produce suboptimal optimizations and may fail to identify appropriate performance bottlenecks. In contrast, \tool adopts a multi-level optimization process based on the MoP architecture to identify suitable optimization targets and generate more efficient programs.


\subsection{Code Efficiency Benchmarks} 
Performance engineering has been studied for many years, and several early benchmarks~\cite{zapa09accuracy,laaber20dynamic,traini23towards} were designed to rigorously evaluate the execution time of programs and software systems. However, these benchmarks are not well suited to evaluating Python programs due to Python's dynamic nature. More recently, several benchmarks have been proposed to evaluate optimization approaches for high-level languages.

\effibench~\cite{effibench} is the first benchmark designed to evaluate the time and memory efficiency of code generation. It selects efficiency-critical problems tagged with ``LeetCode'' and uses GPT-3.5 to generate test cases with different input sizes and data distributions. Mercury~\cite{du24mercury} is another benchmark for evaluating the efficiency of Python programs. It contains 1,889 Python tasks, each with reference solutions that serve as real-world efficiency baselines, enabling a comprehensive analysis of runtime behavior. \coffe~\cite{peng2025coffe} is a benchmark for evaluating both function-level and file-level code generation using LLM-generated stress test cases. By incorporating stress testing, it provides more reliable efficiency evaluation results. In this work, we use \effibench and \coffe to evaluate \tool, as they together provide a comprehensive and reliable basis for assessing code optimization performance.

\section{Conclusion}\label{sec:conclusion}

This paper presents \tool, a multi-level code optimization approach built on a novel Mixture-of-Prompts architecture. \tool performs optimization at four levels to provide comprehensive performance improvement for input programs. At each level, \tool identifies suitable optimization targets, including algorithms and time-critical code structures, and routes them to the appropriate optimization strategies. To support this process, \tool mines optimization strategies from existing code competition datasets and combines them with shared strategies that leverage LLMs' reasoning capabilities. Experimental results show that \tool consistently outperforms existing approaches on both human-written and LLM-generated code, demonstrating the effectiveness of its MoP architecture and multi-level optimization design.

\section{Data Availability}
The code and data in this paper are publicly available at \url{https://doi.org/10.5281/zenodo.19335907}.

\section*{Acknowledgment}

This work is supported by the OUB Professorial Chair Fund awarded by the Singapore Management University, the National Natural Science Foundation of China (No. 62302437), and the Yongjiang Talent Program (No. 2023A-402-G).


\bibliographystyle{ACM-Reference-Format} 
\bibliography{ref}


\end{document}